\documentclass[]{pasj02} 
\usepackage[switch,mathlines]{lineno} 
\usepackage{physics}

\renewcommand{\textcolor}[2]{#2}

\jyear{2024}
\Received{}
\Accepted{}

\begin{document} 

\title{ADF22-WEB: Massive galaxy formation shaped by cosmic web filaments in the $z = 3.1$ proto-cluster core}
\author{
Hideki~Umehata$^{1,2}$,
Ian~Smail$^3$,
Charles~C.~Steidel$^4$,
Bret~D.~Lehmer$^5$,
Kouichiro~Nakanishi$^{6,7}$,
Bunyo~Hatsukade$^{6,7}$,
Mariko~Kubo$^8$,
Norika~Okauchi$^{2}$,
Yoichi~Tamura$^{2}$,
Masato~Hagimoto$^{2}$
}%

\altaffiltext{1}{Institute for Advanced Research, Nagoya University, Furocho, Chikusa, Nagoya 464-8602, Japan}
\altaffiltext{2}{Department of Physics, Graduate School of Science, Nagoya University, Furocho, Chikusa, Nagoya 464-8602, Japan}
\email{umehata@a.phys.nagoya-u.ac.jp}

\altaffiltext{3}{Centre for Extragalactic Astronomy, Department of Physics, Durham University, South Road, Durham DH1 3LE, UK}

\altaffiltext{4}{Cahill Center for Astronomy and Astrophysics, California Institute of Technology, MS249-17, Pasadena, CA91125, USA}

\altaffiltext{5}{Department of Physics, University of Arkansas, 226 Physics Building, 825 West Dickson Street, Fayetteville, AR 72701, USA}

\altaffiltext{6}{National Astronomical Observatory of Japan, 2-21-1 Osawa, Mitaka, Tokyo 181-8588, Japan}
\altaffiltext{7}{Department of Astronomical Science, The Graduate University for Advanced Studies, SOKENDAI, 2-21-1 Osawa, Mitaka, Tokyo
181-8588, Japan}

\altaffiltext{8}{School of Science, Kwansei Gakuin University, Sanda, Hyogo 669-1337, Japan}


\maketitle

\begin{abstract}
\textcolor{blue}{We present a comprehensive census of dust, molecular gas, and galaxy structure in 18 dusty star-forming galaxies (DSFGs) at $z=3.09$ embedded in Ly$\alpha$-traced cosmic web filaments in the core of the SSA22 proto-cluster, based on multi-band ALMA observations, JWST imaging, and CO(1--0) measurements with JVLA. 
Using up to six-band ALMA photometry combined with Herschel/SPIRE data, we construct rest-frame far-infrared spectral energy distributions. The DSFGs span two orders of magnitude in far-infrared luminosity ($L_{\rm FIR}\sim10^{11}$--$10^{13}\,L_\odot$), \textit{with median values of $\log(L_{\rm FIR}/L_\odot)=11.56^{+0.32}_{-0.37}$} and $T_{\rm dust}=25^{+6}_{-3}\,{\rm K}$.
For eight DSFGs with high-resolution ($\sim1$~kpc) JWST and ALMA imaging, we measure stellar and dust surface densities and find a positive correlation, suggesting that structural compaction of the stellar component proceeds together with that of the interstellar medium. 
We report 12 detections of CO(1--0), 18 of CO(3--2), four each of CO(8--7) and CO(9--8), and one of CO(12--11). 
The median brightness temperature ratio between CO(3--2) and CO(1--0) is $r_{31}=0.66_{-0.04}^{+0.05}$, consistent with field galaxies at similar redshifts. 
The high-$J$ CO lines show relatively low excitation, and the CO(8--7)/CO(3--2) ratio correlates with star-formation rate surface density, suggesting that molecular gas excitation is primarily regulated by star formation.
Lower-mass galaxies show excess molecular gas fractions and depletion times relative to field scaling relations.
By combining quiescent galaxy samples in the same field, we identify a sequence from gas-rich systems with low stellar surface density to gas-poor, compact systems with high stellar surface density. 
These results suggest that massive galaxy evolution in dense environments is governed by a baryon cycle linking gas supply, star formation, and structural transformation within the cosmic web environment.}
\end{abstract}



\section{Introduction}


The central regions of galaxy clusters in the local Universe are dominated by massive early-type galaxies that show little or no ongoing star formation \citep{1980ApJ...236..351D}. In contrast, at $z\gtrsim2$, proto-clusters—overdense regions thought to be the progenitors of present-day clusters—are often populated by dusty star-forming galaxies (DSFGs), which exhibit intense star-formation rates and rapid stellar mass growth (e.g., \cite{2003Natur.425..264S,2003ApJ...585...57C,2009Natur.459...61T,2009ApJ...694.1517D,2012Natur.486..233W,2014MNRAS.440.3462U,2014A&A...570A..55D,2015ApJ...815L...8U,2016ApJ...824...36C,2016ApJ...827...34O,2016ApJ...828...56W,2017ApJ...842...55L,2018Natur.556..469M,2025ApJ...983...69W,2025A&A...696A..33P}). While these systems are widely considered plausible progenitors of the massive elliptical galaxies observed in cluster cores today \citep{2006MNRAS.366..499D}, the physical mechanisms that drive their rapid mass growth and subsequent quenching remain poorly understood.

A key aspect of this transformation is the cold interstellar medium (ISM), composed of dust and molecular gas. A comprehensive understanding of this component across different evolutionary stages—from the early and peak phases of active star formation to quiescent systems—is essential for unveiling the overall picture of star formation and galaxy assembly during the most active epoch of massive galaxies in present-day clusters \citep{2002PhR...369..111B,2014PhR...541...45C,2016A&ARv..24...14O,2022Univ....8..554A}.
While the Atacama Large Millimeter/submillimeter Array (ALMA) has revolutionized our ability to probe the cold ISM at high redshift, our understanding remains limited. One of the primary limitations is the sparse sampling of dust spectral energy distributions (SEDs), which directly affects our ability to constrain dust mass, temperature, and radiation field properties in a consistent manner.
Although several studies have constructed valuable statistical samples of DSFG SEDs \textcolor{magenta}{covering the far-infrared-to-submillimeter regime} \textcolor{blue}{(e.g., \cite{2014MNRAS.438.1267S,2020MNRAS.494.3828D,2020ApJ...902...78R, 2021ApJ...919...30D, 2025MNRAS.540.1560B})}, the number of submillimeter photometric data points free from source confusion—particularly those obtained with ALMA—is often limited. In addition, uncertainties in redshift, in cases where spectroscopic measurements are unavailable, further propagate into uncertainties in derived quantities such as dust temperature and far-infrared luminosity.
\textcolor{blue}{In lensed DSFGs, differential lensing effects may further introduce systematic uncertainties.}
Moreover, these studies tend to be biased toward the brightest submillimeter sources, leaving the pre- and post-peak phases of starburst activity largely unexplored.

The molecular gas content—the direct fuel of star formation—remains incompletely characterized \textcolor{blue}{across all environments}, including in proto-cluster environments. One major limitation is the scarcity of CO(1--0) observations \textcolor{blue}{(e.g., \cite{2011MNRAS.412.1913I,2018ApJ...867L..29W,2021ApJ...913..110C,2023ApJ...945..128F,2025A&A...700A.278R, 2026ApJS..282...40P})}, which provide the most robust tracer of the total molecular gas reservoir. Instead, many studies rely on higher-$J$ CO transitions (e.g., \textcolor{blue}{\cite{2013MNRAS.429.3047B, 2017ApJ...842...55L,2019PASJ...71...40T,2021MNRAS.501.3926B, 2025A&A...701A.234Z})}, which require uncertain excitation corrections to infer the total gas mass.
In addition, the treatment of the CO-to-H$_2$ conversion factor ($\alpha_{\rm CO}$) remains a major source of systematic uncertainty. While DSFGs are intensely star-forming systems, they are often found to lie on or near the star-forming main sequence (e.g., \cite{2015ApJ...806..110D,2020MNRAS.494.3828D,2025A&A...699A.324H}). Together with the difficulty of measuring gas-phase metallicities \textcolor{blue}{(Taylor et al in preparation)}, 
this has led to the adoption of both starburst-like ($\alpha_{\rm CO}\sim0.8$) and main-sequence/Milky-Way-like ($\alpha_{\rm CO}\sim4$) conversion factors in the literature (e.g., \cite{2011MNRAS.412.1913I,2013MNRAS.429.3047B,2018ApJ...867L..29W,2019ApJ...882..140B,2020ApJ...896L..21R,2020A&A...635A.119C,2021MNRAS.501.3926B,2024ApJ...961..226L,2025PASJ...77..432U,2025A&A...701A.234Z}), introducing significant uncertainties in the derived gas masses and related quantities.
Moreover, high-$J$ CO transitions, which trace dense and warm gas more directly linked to star formation, have begun to be explored, but current studies remain limited \textcolor{blue}{(e.g., \cite{2013MNRAS.429.3047B,2021MNRAS.501.3926B,2025MNRAS.536.1149T,2026arXiv260223521T})}.
This situation hinders a consistent understanding of the molecular gas properties and star-formation processes in protoclusters.

Another major uncertainty lies in the connection between the ISM and stellar structure. Observations suggest that a significant fraction of DSFGs exhibit disk-like morphologies (e.g., \cite{2022ApJ...939L...7C,2024A&A...691A.299G,2025ApJ...978..165H,2026ApJ...997...79U}), whereas their presumed descendants, quiescent galaxies (QGs), are generally characterized as compact systems \citep{2014ApJ...792L...6V}. While several studies have attempted to link these populations based on the sizes of dust-emitting regions \textcolor{blue}{(e.g., \cite{2015ApJ...810..133I,2015ApJ...799...81S,2015ApJ...799..194C,2019MNRAS.490.4956G, 2019ApJ...876..130H,2020A&A...635A.119C,2022ApJ...939L...7C})}, the physical pathway of this structural transformation remains unclear. 
In particular, it is still debated whether this transition is driven primarily by gas consumption, compaction, or environmental processes\textcolor{blue}{, each of which is expected to leave distinct signatures in the spatial distributions of the ISM and stellar components. In this context, ``compaction'' refers to the buildup of centrally concentrated gas and stellar components, leading to enhanced stellar surface densities and compact galaxy structures.}
This uncertainty highlights the need for spatially resolved measurements of both the ISM and stellar components.

Finally, recent observations have begun to directly reveal large-scale filamentary structures of the cosmic web (e.g., \cite{2019Sci...366...97U,2019PASJ...71L...2K,2021A&A...649A..78D,2017ApJ...837...71C,2025A&A...696A..33P}), opening a new avenue to investigate their role in regulating gas accretion and galaxy evolution \citep{2006MNRAS.368....2D,2022MNRAS.514.5429C}. However, the interplay between filamentary gas supply, star formation, and quenching remains poorly constrained observationally.
This motivates a joint view that connects large-scale structure with the baryon cycle in galaxies.

In this paper, \textcolor{blue}{we present a comprehensive study of DSFGs in the SSA22 proto-cluster at $z=3.09$ \citep{1998ApJ...492..428S}, focusing on the ALMA Deep Field in SSA22 (ADF22), a deep multi-wavelength survey field in the protocluster core where cosmic web filaments have been directly mapped through Ly$\alpha$ emission and a close association between DSFGs and QGs has been established, and discuss their evolutionary connection to QGs} (e.g., \cite{2019Sci...366...97U,2021ApJ...919....6K,2022ApJ...935...89K,2025PASJ...77..432U,2025ApJ...985L...8U,2025A&A...699A.324H,2026ApJ...997...79U}).
We analyze 18 DSFGs at $z=3.09$ by combining multi-band ALMA continuum data, CO(3--2) observations available for all sources (\cite{2019Sci...366...97U}), CO(1--0) and CO(8--7) measurements for subsets of the sample obtained with ALMA and the Karl G. Jansky Very Large Array (VLA) (\cite{2025ApJ...985L...8U}), and high-resolution dust continuum and near-infrared imaging from ALMA and JWST (\cite{2026ApJ...997...79U}). This dataset enables us to characterize the dust properties, molecular gas content and excitation, and structural properties of galaxies across a wide range of environments within the proto-cluster, in direct comparison with QGs, thereby linking galaxy-scale evolution to the surrounding cosmic web.

This paper is organized as follows. In \S2 we describe the observations and data reduction. In \S3 we present the analysis of the dust continuum data and derive dust properties. In \S4 we analyze the molecular gas properties. In \S5 we combine these results to investigate the scaling relations, as well as the evolution and transformation of massive galaxies.
We adopt a flat $\Lambda$CDM cosmology with $H_0=70$\,km\,s$^{-1}$\,Mpc$^{-1}$, $\Omega_{\rm m}=0.30$, and $\Omega_\Lambda=0.70$, which corresponds to a physical scale of 7.63\,kpc\,arcsec$^{-1}$ at $z=3.09$.


\section{Observation and data reduction}

\begin{figure}
  \begin{center}
\includegraphics[width=8.5cm]
{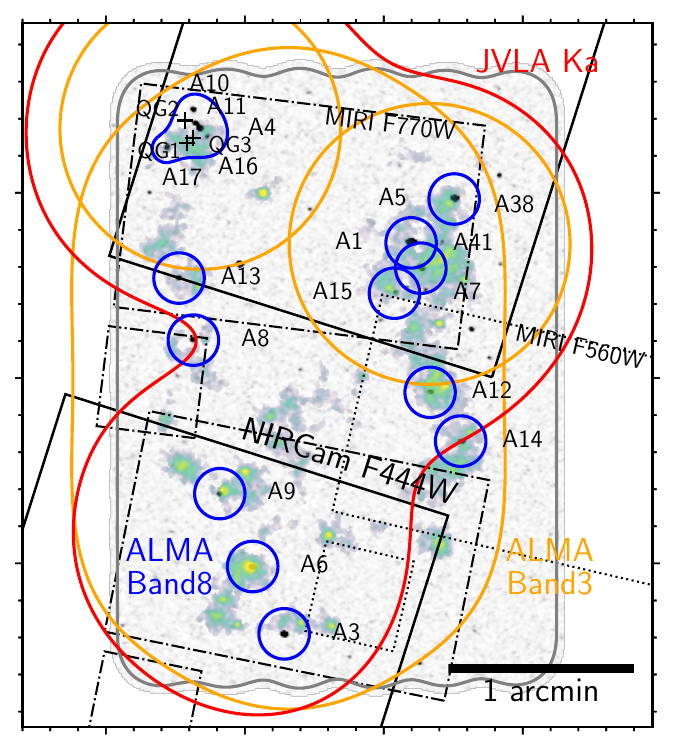}
  \end{center}
  \caption{%
ALMA 1.17\,mm map of the ADF22 field in the $z=3.1$ SSA22 proto-cluster core, shown in grayscale.
Ly$\alpha$ filaments are overlaid as colored regions for comparison (\cite{2019Sci...366...97U}).
The representative coverages of ALMA Band~3, Band~8, and JVLA Ka-band are shown as orange, blue, and red contours, respectively, corresponding to regions with a primary beam response $>30\%$. 
For ALMA Band~3, two additional deep pointings are available besides the wide mosaic. 
The 18 DSFGs \textcolor{blue}{and 3 QGs} at $z_{\rm spec}\simeq3.09$ are labeled (\cite{2015ApJ...815L...8U,2019Sci...366...97U,2025A&A...699A.324H}). 
%
The ADF22 DSFGs, many of which lie along the filaments, are largely encompassed by these coverages as well as JWST/NIRCam \textcolor{blue}{and/or MIRI} imaging (\cite{2026ApJ...997...79U}), enabling a joint characterization of molecular gas, dust spectral energy distributions, and stellar properties within the filamentary structure.
{Alt text: A map indicating observation regions.} 
}%
  \label{fig:fov}
\end{figure}

\subsection{Data Overview}

\begin{figure*}
  \begin{center}
\includegraphics[width=16.5cm]
{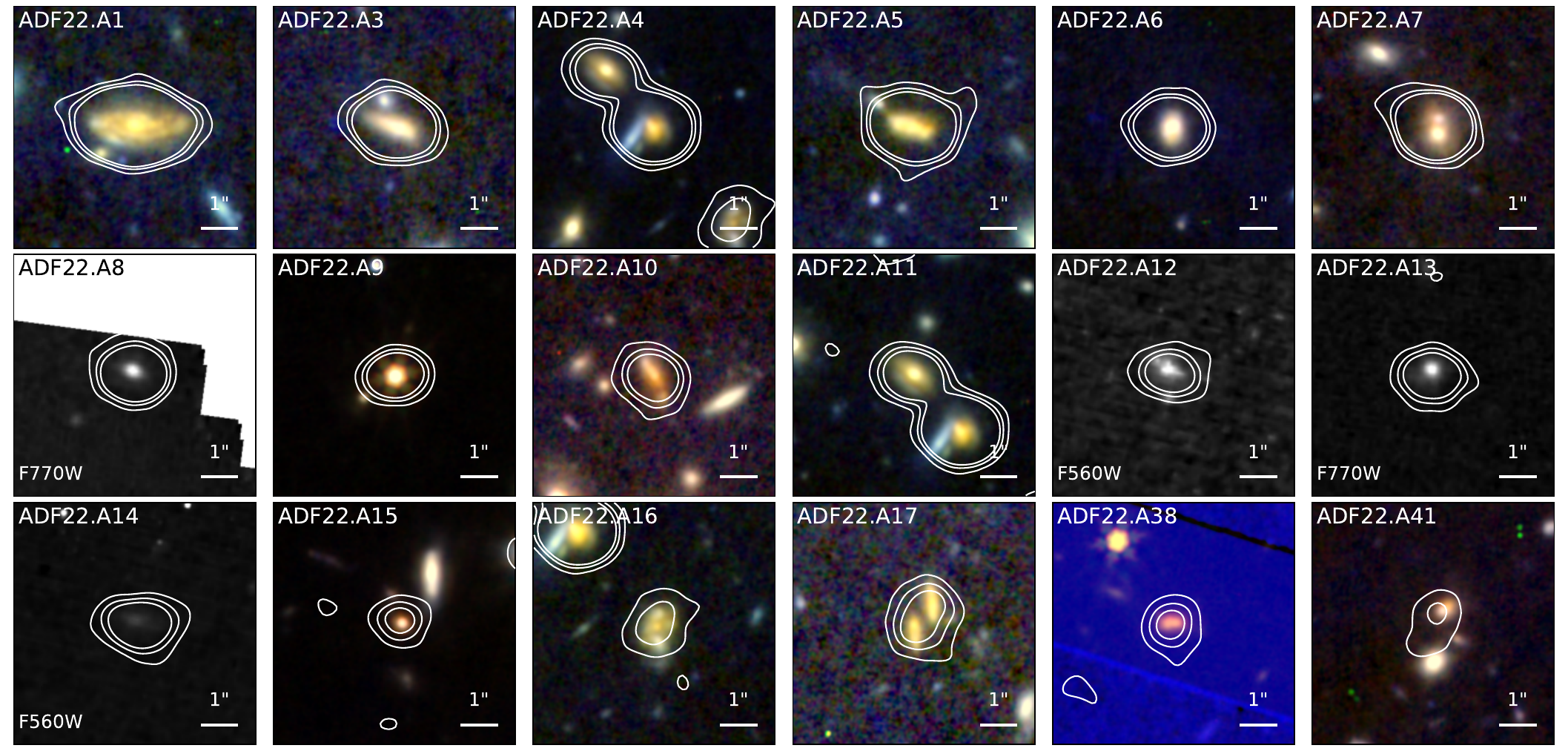}
  \end{center}
  \caption{%
ALMA and JWST views of the 18 DSFGs in the ADF22 field at $z_{\rm spec}\simeq3.09$. 
Background color images show pseudo-color composites of the JWST NIRCam
F200W, F356W, and F444W data (\cite{2026ApJ...997...79U}).
For sources not covered by the NIRCam imaging, single-band MIRI images are
shown in grayscale \textcolor{blue}{as labeled}.
Contours indicate the ALMA 1.17\,mm emission at $3$, $6$, and $9\sigma$
(\cite{2025A&A...699A.324H}).
\textcolor{blue}{Each panel is centered on the source indicated by the corresponding ID label.}
The JWST images reveal diverse stellar morphologies among the DSFGs,
ranging from compact sources to extended and multiple-component systems.
Each panel is $7''\times7''$, corresponding to $\sim54$\,kpc at $z=3.09$.
{Alt text: Images of target galaxies.} 
}%
  \label{fig:finding_chart}
\end{figure*}

Fig.~\ref{fig:fov} shows the deep 1.17\,mm image of ADF22 obtained by combining ALMA Cycle~2 and Cycle~5 data. Details of the observations and source catalogs are presented in \citet{2025A&A...699A.324H}.
The field contains 16 bright DSFGs at $z_{\rm spec}\simeq3.09$ (ADF22.A1--ADF22.A17; \cite{2015ApJ...815L...8U}; \yearcite{2017ApJ...835...98U}; \yearcite{2018PASJ...70...65U}; \yearcite{2019Sci...366...97U}). In addition, two fainter DSFGs are newly identified based on the sensitive CO(3--2) survey presented in this paper (see also \cite{2026ApJ...997...79U}). 
These DSFGs have been subject to extensive multi-wavelength follow-up observations with facilities including ALMA, JVLA, and JWST. The corresponding fields of view are indicated in Fig.~\ref{fig:fov}. Details of the individual observations are described in the following subsections (see also \cite{2025PASJ...77..432U}; \yearcite{2025ApJ...985L...8U}; \yearcite{2026ApJ...997...79U}).

\subsection{JVLA Ka-band}

The CO(1--0) line at $z\simeq3.09$ falls within the Ka-band of JVLA \citep{2011ApJ...739L...1P}. Observations of the ADF22 field were conducted between winter 2016 and summer 2021 under two programs (16A--357 and 21A--346; PI: H.~Umehata).
We utilized the Wideband Interferometric Digital Architecture (WIDAR) correlator with 8-bit samplers to obtain dual-polarization data with a spectral resolution of 2\,MHz. Two frequency setups with slightly offset intermediate frequencies (IFs) were adopted to mitigate gaps between sub-bands. The Ka-band receiver covered two basebands, each with a bandwidth of 1\,GHz, centered at 32.748\,GHz (or 32.766\,GHz) and 28.462\,GHz (or 28.480\,GHz). These configurations provided continuous frequency coverage of $32.67$--$33.70$\,GHz and $27.67$--$28.70$\,GHz, corresponding to CO(1--0) at $z=3.02$--$3.17$ and $z=2.42$--$2.53$, respectively. The frequency coverage fully encompasses the CO(1--0) emission from galaxies in ADF22 at $z\simeq3.09$.
Each target was observed with on-source integration times of $4$--$5$\,minutes per \textcolor{blue}{scan}, interleaved with observations of the nearby quasar J2218--0335 for complex gain calibration. Each track also included observations of 3C48 for absolute flux calibration.

Part of the ADF22 field was previously observed by \citet{2016ApJ...827...18S} targeting CO(1--0) at $z=3.09$, motivated by the presence of the DSFG ADF22.A3, which corresponds to the known source SMM\,J22174$+$0015 \citep{2005MNRAS.359.1165G,2013MNRAS.429.3047B}. We incorporate archival data from program 15B--329, which are compatible with our JVLA observations and improve the overall sensitivity.
As a result, the majority of the ADF22 field is covered by five  pointings (Fig.~\ref{fig:fov}; see also Appendix~\ref{app:jvla_obs}). Multiple array configurations (D, C, and CnB) were used, and the integration times vary among pointings, resulting in inhomogeneous sensitivity. 

The raw visibilities were calibrated and flagged using the CASA pipeline within the Common Astronomy Software Applications (\textsc{casa}) package (version 6.1.2; \cite{2007ASPC..376..127M,2022PASP..134k4501C}). All calibrated $uv$ data were concatenated into a single dataset. Imaging was performed using the \texttt{tclean} task in \textsc{casa} with Briggs weighting (robust = 2.0). 
Given the inhomogeneous $uv$ coverage, imaging was carried out separately for the three main fields, merging only a close pair. We also constructed a combined data cube using all available data to assess the overall spatial coverage (Fig.~\ref{fig:fov}). A $uv$ taper was applied to achieve a more homogeneous synthesized beam among the fields (A1A5A7A15, A4A10A11A16A17, and A3A6A9). During cleaning, auto-masking was adopted with parameters (\texttt{noisethreshold}=4.0, \texttt{lownoisethreshold}=1.5), and cleaning was performed down to the $2\sigma$ level. The resulting r.m.s. noise is $30$--$40\,\mu\mathrm{Jy\,beam^{-1}}$ per 100\,km\,s$^{-1}$ at $28.0$--$28.4$\,GHz.
The resulting synthesized beam is $\sim 3.3'' \times 2.6''$ (Table~\ref{table:obs}).

\subsection{ALMA Band~3}

ALMA Band~3 observations were conducted to target the redshifted CO(3--2) line and 3\,mm dust continuum emission from proto-cluster galaxies in the ADF22 field. 
Two types of observations were carried out. 
First, a 13-pointing mosaic was obtained to encompass the majority of the ALMA Deep Field, as briefly reported in \citet{2019Sci...366...97U}. 
Second, deeper pointed observations were conducted toward the AzTEC1 group (ADF22.A1, A5, A7, A15) and the AzTEC14 group (ADF22.A4, A10, A11, A16, A17). 

\subsubsection{ADF22 mosaic}

The observations were conducted during two periods, from 2016 July 21 to August 19 and from 2017 May 7 to July 4, as part of Cycle~3 and Cycle~4 programs (project IDs: 2015.1.00212.S and 2016.1.00543.S; PI: H.~Umehata). 
A total of 19 execution blocks (EBs) were obtained. 
The baseline lengths range from 15\,m to 2647\,m, corresponding to a representative array configuration of C40--5. 
A 13-pointing mosaic with Nyquist sampling was obtained in each EB, centered at ($\alpha$, $\delta$) = (22$^{\rm h}$17$^{\rm m}$35.0$^{\rm s}$, $+00^\circ17^\prime00.0^{\prime\prime}$; ICRS). 
The mosaic covers the majority of the 1.1\,mm image in ADF22A with a primary-beam response $>50\%$ (Fig.~\ref{fig:fov}). 
The correlator was configured in Frequency Division Mode (FDM), with four spectral windows of 1.875\,GHz bandwidth each and a spectral resolution of 977\,kHz. 
The spectral windows were centered at 85.04, 86.87, 97.03, and 98.87\,GHz, enabling coverage of CO(3--2) at $z=2.47$--2.60 and $z=2.94$--3.11. 
The observations were conducted with 36--44 12\,m antennas under favorable Band~3 conditions, with precipitable water vapor (PWV) of 0.4--2.5\,mm. 
The total on-source integration time is 13.9\,hr. 

The ALMA data were reduced \textcolor{blue}{using the \textsc{casa} and ALMA Pipeline} (versions 4.7.0, 4.7.2, 5.1.0, and 6.1.0). 
All $uv$ data from different EBs were concatenated using the \texttt{concat} task in \textsc{casa}, adopting a frequency tolerance of \texttt{freqtol}=`5\,MHz'. 
Imaging was performed using the \texttt{tclean} task in \textsc{casa}. 
A dirty cube was first generated to identify line-free channels, and the visibilities were continuum-subtracted after masking line-emitting channels. 
The continuum-subtracted visibilities were then imaged into spectral cubes. 
Natural weighting was adopted to maximize sensitivity for line detection. 
During imaging, auto-masking (\cite{2020PASP..132b4505K}) was applied with \texttt{noisethreshold}=4.5 and \texttt{lownoisethreshold}=1.5, and cleaning was performed down to the $2\sigma$ level. 
The resulting synthesized beam is $0.94^{\prime\prime} \times 0.80^{\prime\prime}$ (PA=$-55^\circ$) at 84.5\,GHz, corresponding to CO(3--2) at $z\simeq3.09$. 
The achieved r.m.s. noise level is $89\,\mu\mathrm{Jy\,beam^{-1}}$ per 100\,km\,s$^{-1}$ channel at the phase center. 
%

\subsubsection{Deeper mappings}

We conducted deeper observations toward two dense galaxy groups, AzTEC1 and AzTEC14, originally identified as single sources in the AzTEC/ASTE survey (\cite{2009Natur.459...61T,2014MNRAS.440.3462U}). 
%
The CO(3--2) line ($\nu_{\rm rest}=345.796\,\mathrm{GHz}$) was observed toward the AzTEC1 group as part of project 2022.1.00680.S (PI: Umehata). 
The phase center was set to ($\alpha$, $\delta$) = (22$^{\rm h}$17$^{\rm m}$32.0$^{\rm s}$, $+00^\circ17^\prime43.5^{\prime\prime}$), covering ADF22.A1, A5, A7, and A15. 
The observations were carried out in 27 executions during Cycle~9 (2022 December to 2023 March) with array configurations C--3 to C--5. 
The total on-source integration time is 21\,hr. 
The correlator setup is similar to that of the mosaic observations, with four spectral windows centered at 85.03, 86.86, 97.03, and 98.86\,GHz. 
Data reduction was performed using \textsc{casa} version 6.5.4. 
Imaging was carried out with \texttt{tclean} using natural weighting and auto-masking, cleaned down to $2\sigma$. 
The synthesized beam is $1.49^{\prime\prime} \times 1.35^{\prime\prime}$ (PA=$69^\circ$), and the typical r.m.s. noise level at the phase center is $30\,\mu\mathrm{Jy\,beam^{-1}}$ per 100\,km\,s$^{-1}$ channel at $\sim85$\,GHz. 
The AzTEC14 observations were obtained as part of project 2023.1.01206.S (PI: Umehata), as described in \citet{2025ApJ...985L...8U}. 
\textcolor{blue}{The typical r.m.s. noise is approximately $50\,\mu\mathrm{Jy\,beam^{-1}}$ per 80\,km\,s$^{-1}$ channel, with a synthesized beam of $1.94^{\prime\prime} \times 1.56^{\prime\prime}$ (PA=$84^\circ$)}.

\begin{figure*}
  \begin{center}
\includegraphics[width=17cm]
{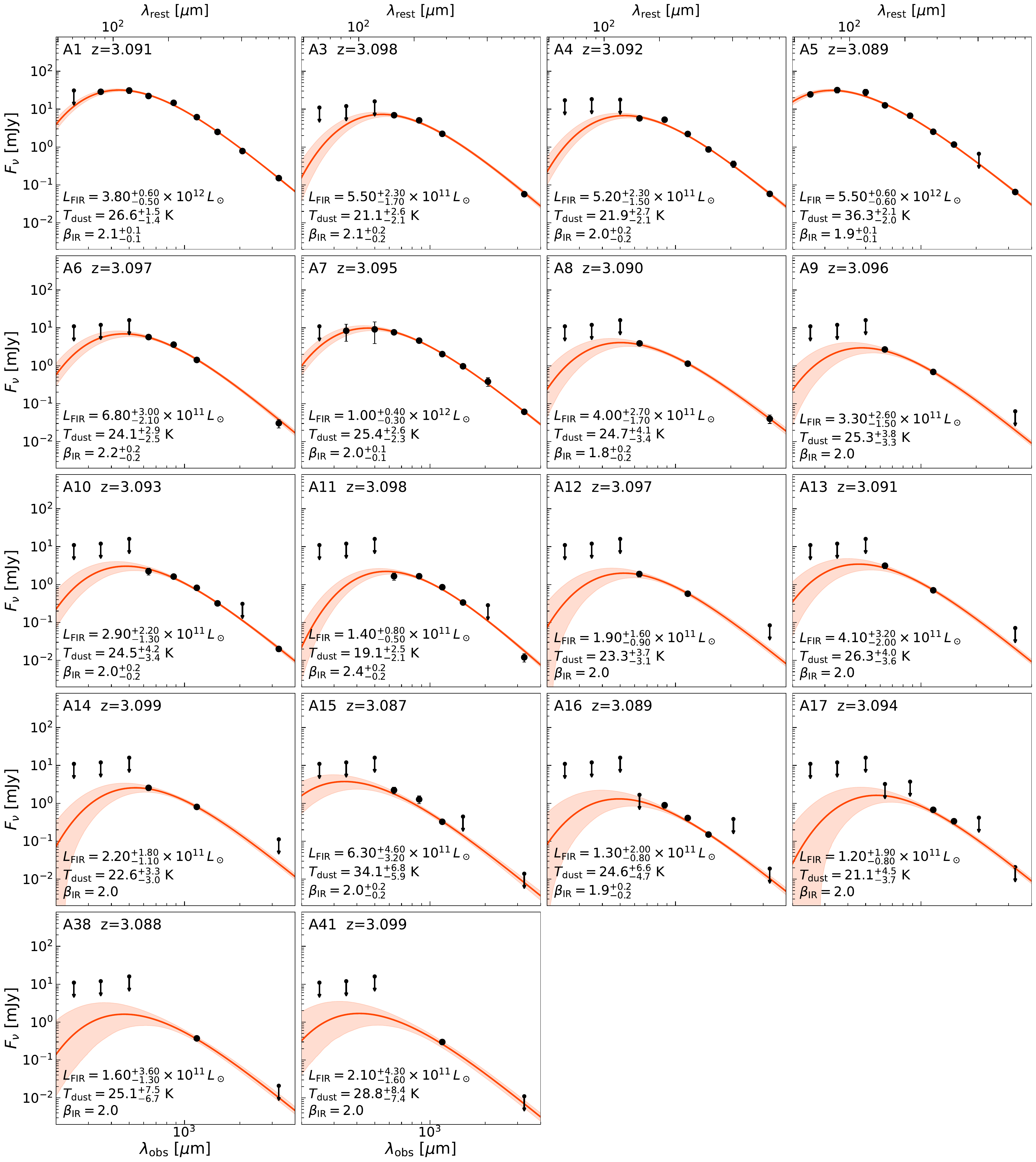}
  \end{center}
  \caption{
  Spectral energy distributions (SEDs) of the 18 DSFGs at $z\simeq3.09$. Each panel shows the observed-frame flux density $F_\nu$ (mJy) as a function of the observed wavelength, $\lambda_{\rm obs}$ ($\mu$m).
Filled circles with error bars indicate detections, while arrows denote 3$\sigma$ upper limits.
The solid curve represents the maximum-likelihood SED obtained using the {\sc mercurius} fitting routine
(\cite{2022MNRAS.515.1751W}).
The shaded region shows the 16th--84th percentile range of the posterior distribution at each wavelength.
The inferred dust temperature and dust emissivity index are indicated in each panel.
{Alt text: spectral energy distributions.} 
}%
  \label{fig:sed}
\end{figure*}

\subsection{ALMA Band~6}

\textcolor{blue}{ALMA Band~6 observations targeting the CO(8--7) and CO(9--8)
emission lines from ADF22.A1, A4, A6, and A7 were carried out as part of project 2021.1.00511.S (PI: H.~Umehata).
The observations were conducted between May and August 2022,
with total on-source integration times of 32--33\,min for the CO(8--7) setup and 44--49\,min for the CO(9--8) setup.
The correlator was configured with four spectral windows,
each with a bandwidth of 1.875\,GHz, covering the redshifted
CO(8--7) and CO(9--8) lines at observed frequencies of
225.4\,GHz and 253.6\,GHz, respectively, for $z \sim 3.09$.
Data calibration was performed using \textsc{casa}
(version 6.2.1). Imaging was carried out with
\texttt{tclean} using natural weighting and auto-masking,
and cleaned down to the $2\sigma$ level.
The CO(8--7) cubes typically achieve a synthesized beam size of
$0.98^{\prime\prime} \times 0.86^{\prime\prime}$
(PA=$61^\circ$), with an r.m.s.\ noise level of
$0.16\,\mathrm{mJy\,beam^{-1}}$
per 80\,km\,s$^{-1}$ channel at 225.4\,GHz.
For the CO(9--8) data, the A1 and A7 fields have synthesized
beams of
$0.92^{\prime\prime} \times 0.81^{\prime\prime}$
(PA=$-84^\circ$),
with an r.m.s.\ noise level of
$0.11\,\mathrm{mJy\,beam^{-1}}$
per 80\,km\,s$^{-1}$ channel at 253.7\,GHz.
For the A4 and A6 fields, we applied $uv$ tapering to obtain
a synthesized beam of
$0.69^{\prime\prime} \times 0.62^{\prime\prime}$
(PA=$58^\circ$),
resulting in an r.m.s.\ noise level of
$0.17\,\mathrm{mJy\,beam^{-1}}$
per 80\,km\,s$^{-1}$ channel at the phase center.}

\subsection{ALMA Continuum Data}

\subsubsection{ALMA Band~3 continuum}

The ADF22 mosaic Band~3 data were also used to create a 3.26\,mm (91.9\,GHz) continuum map using the \texttt{tclean} task in \textsc{casa}, by combining channels free from CO(3--2) emission at $z\simeq3.09$. 
The resulting synthesized beam is $0.87^{\prime\prime} \times 0.74^{\prime\prime}$ (PA=$-59^\circ$), and the r.m.s.\ noise at the phase center is $6\,\mu\mathrm{Jy\,beam^{-1}}$. 
This dataset covers all 16 brightest DSFGs and was used when a $z\simeq3.09$ DSFG was not included in the deeper datasets described below.

In addition, deeper Band~3 observations of the AzTEC1 and AzTEC14 fields were used to create 3.26\,mm continuum maps using line-free channels. 
The resulting synthesized beams are $1.35^{\prime\prime} \times 1.24^{\prime\prime}$ (PA=$70^\circ$) and $1.69^{\prime\prime} \times 1.39^{\prime\prime}$ (PA=$78^\circ$), respectively. 
The achieved r.m.s.\ noise levels are $1.9$ and $2.9\,\mu\mathrm{Jy\,beam^{-1}}$. 
The AzTEC1 image was used to measure the flux densities of A1, A5, A7, and A15, while the AzTEC14 image was used for A4, A10, A11, A16, and A17.

\subsubsection{ALMA Band~4 continuum}

We make use of the ALMA Band~4 continuum data presented in \citet{2022ApJ...930...32C}, which targeted a subset of SCUBA-2–selected 850\,$\mu$m sources as part of project 2019.1.00313.S (PI: C.~Casey), covering part of the AzTEC1 and AzTEC14 regions. 
The calibrated measurement sets were generated by \textcolor{blue}{the East Asian ALMA Regional Center} using the standard pipeline.
Imaging was performed with \texttt{tclean} using natural weighting and auto-masking, and cleaned down to the $2\sigma$ level using line-free channels. 
The representative frequency is 145.0071\,GHz (2.07\,mm). 
The synthesized beam is $1.86^{\prime\prime} \times 1.57^{\prime\prime}$ (PA=$-60^\circ$). 
The achieved 1$\sigma$ sensitivities are $62$ and $64\,\mu\mathrm{Jy\,beam^{-1}}$ for the AzTEC1 and AzTEC14 fields, respectively.

\subsubsection{ALMA Band~5 continuum}

ALMA Band~5 observations were obtained as part of project 2021.1.00511.S (PI: H.~Umehata). 
The dataset consists of two pointings centered on ADF22.A1 and ADF22.A4, covering A1, A4, A5, A10, A11, A15, A16, and A17. 
The observations were conducted in August 2022 using the C-4/C-5 configurations, with total on-source integration times of 72\,min (ADF22.A1) and 96\,min (ADF22.A4).
The correlator was configured with four spectral windows (1.875\,GHz each) centered at 191.35, 193.22, 203.35, and 205.03\,GHz. 
Data calibration was performed using \textsc{casa} (version 6.2.1). 
Continuum imaging was carried out with \texttt{tclean} using natural weighting and auto-masking, and cleaned down to the $2\sigma$ level.
For the ADF22.A1 field, a $uv$ taper of $0.7^{\prime\prime}$ was applied, resulting in a synthesized beam of $1.05^{\prime\prime} \times 0.88^{\prime\prime}$ (PA=$64^\circ$) and an r.m.s.\ noise of $23\,\mu\mathrm{Jy\,beam^{-1}}$ at 198.16\,GHz (1.51\,mm). 
For the ADF22.A4 field, a $uv$ taper of $0.5^{\prime\prime}$ was applied, yielding a synthesized beam of $0.97^{\prime\prime} \times 0.77^{\prime\prime}$ (PA=$90^\circ$) and an r.m.s.\ noise of $10\,\mu\mathrm{Jy\,beam^{-1}}$.

\subsubsection{ALMA Band~6 continuum}

All 18 DSFGs at $z\simeq3.09$ were originally identified through ALMA Band~6 observations \citep{2015ApJ...815L...8U,2017ApJ...835...98U,2018PASJ...70...65U,2020A&A...640L...8U,2025A&A...699A.324H}. 
We adopt the latest source catalog compiled by \citet{2025A&A...699A.324H}. 
The Band~6 continuum images at 1.17\,mm typically achieve an r.m.s.\ noise level of $\sigma=25\,\mu\mathrm{Jy\,beam^{-1}}$ with a synthesized beam of $1.3^{\prime\prime} \times 1.1^{\prime\prime}$ (PA=$88^\circ$).

\subsubsection{ALMA Band~7 continuum}

Six of the brightest sources (ADF22.A1, ADF22.A3, ADF22.A4, ADF22.A5, ADF22.A6, and ADF22.A7) were observed at 870\,$\mu$m with ALMA Band~7 in single-pointing mode \citep{2026ApJ...997...79U}. 
A $uv$ taper of $0.6^{\prime\prime}$ was applied to avoid resolving out extended emission. 
The resulting synthesized beam is $0.92^{\prime\prime} \times 0.82^{\prime\prime}$ (PA=$88^\circ$), with an r.m.s.\ noise of $60\,\mu\mathrm{Jy\,beam^{-1}}$.

\subsubsection{ALMA Band~8 continuum}

As reported in \citet{2025PASJ...77..432U}, ALMA Band~8 observations were obtained as part of Cycle~8 project 2021.1.00041.S (PI: H.~Umehata). 
The observations were conducted in April 2022 using the C43-2 configuration, with on-source integration times of 14--23\,min per pointing.
Data calibration and imaging were performed using standard \textsc{casa} procedures. 
Continuum images were created with \texttt{tclean} using natural weighting and a $uv$ taper of $0.75^{\prime\prime}$ to improve surface-brightness sensitivity. 
The representative frequencies are 470.67 and 470.10\,GHz ($\sim640\,\mu$m), and the synthesized beam is $\sim0.90^{\prime\prime} \times 0.85^{\prime\prime}$ (PA$\approx70^\circ$). 
The typical 1$\sigma$ sensitivity is $0.2$--$0.3\,\mathrm{mJy\,beam^{-1}}$.

\subsection{JWST Images}

The ADF22 field has been observed multiple times with JWST, and the corresponding fields of view are shown in Fig.~\ref{fig:fov}. 
NIRCam imaging was obtained under GO~3547 (PI: Umehata), providing coverage in the F115W, F200W, F356W, and F444W filters (see \cite{2026ApJ...997...79U} for details). 
MIRI F560W imaging was also obtained as part of GO~3547 \textcolor{blue}{(PI: H.~Umehata)}, while MIRI F770W and F2100W observations were carried out under GO~6751 (PI: Umehata). 
Details of the MIRI data will be presented in a forthcoming paper. 
In this work, we use the MIRI images for a subset of the ALMA DSFGs not covered by the four NIRCam bands, in order to complement the rest-frame optical view of their stellar components.

\section{Dust Continuum Views}

\begin{figure}
  \begin{center}
\includegraphics[width=8cm]
{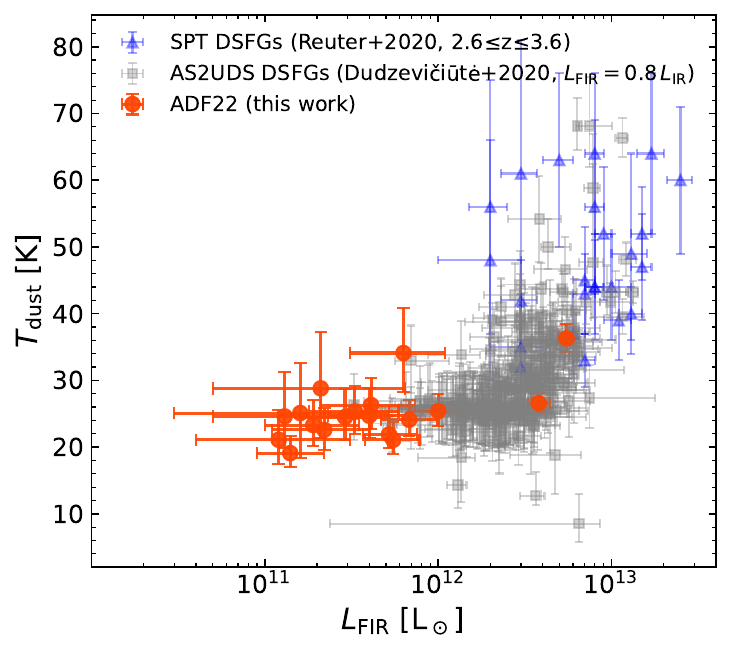}
  \end{center}
\caption{
Relation between the far-infrared luminosity ($L_{\rm FIR}$) and dust
temperature ($T_{\rm dust}$) for the ADF22 DSFGs at $z \simeq 3.09$
(orange circles).
For comparison, SPT-selected DSFGs at $z_{\rm spec}=2.6-3.6$ (\cite{2020ApJ...902...78R}) and AS2UDS DSFGs at $z_{\rm phot}=2.6-3.6$ (\cite{2020MNRAS.494.3828D}) are shown as blue and grey symbols, respectively.
For the AS2UDS sample, $L_{\rm FIR}$ is approximated as $0.8\,L_{\rm IR}$ to match the wavelength range considered.
The ADF22 DSFGs occupy a lower-luminosity and moderately warm dust regime, representative of less extreme dusty star-forming galaxies at $z \sim 3$.
{Alt text: Relation between luminosity and temperature.} 
}%
  \label{fig:fir_td}
\end{figure}

\begin{figure*}
  \begin{center}
\includegraphics[width=16cm]
{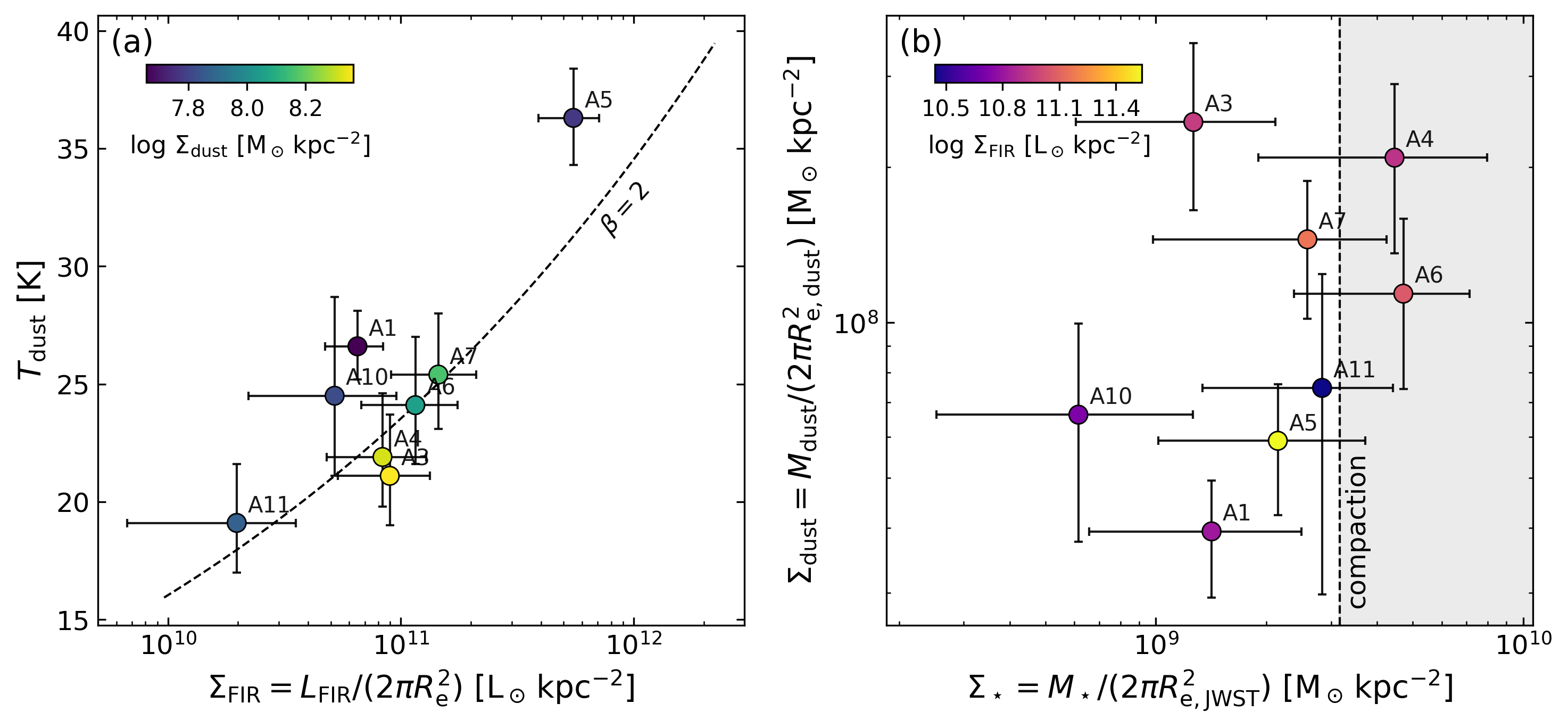}
  \end{center}
  \caption{
Relation between far-infrared surface brightness, dust temperature, stellar mass surface density, and dust mass surface density for the ADF22 DSFGs at $z\simeq3.09$.
 (a) Dust temperature ($T_{\rm dust}$) as a function of far-infrared surface brightness, $\Sigma_{\rm FIR}=L_{\rm FIR}/(2\pi R_{\rm e}^2)$.
The dashed line shows the expected scaling for optically thin dust emission with $\beta=2$, normalized to the median $\Sigma_{\rm dust}$ of the sample. Colors indicate the dust mass surface density, $\Sigma_{\rm dust}$, showing that the offset from the relation depends on $\Sigma_{\rm dust}$.
(b) Dust mass surface density, $\Sigma_{\rm dust}=M_{\rm dust}/(2\pi R_{\rm e,dust}^2)$, versus stellar mass surface density, $\Sigma_\star=M_\star/(2\pi R_{{\rm e}, {\rm JWST}}^2)$.
Colors indicate $\Sigma_{\rm FIR}$. The two surface densities broadly correlate, indicating coupled mass concentration in the stellar component and the cold ISM. The shaded region marks the compact (compaction) regime defined by $\log \Sigma_\star \gtrsim 9.5\ ({\rm M_\odot\,kpc^{-2}})$ (\cite{2017ApJ...840...47B}), where the most compact DSFGs are located.
{Alt text: Diagrams showing temperature and density} 
}%
  \label{fig:sigma_fir_td}
\end{figure*}

\subsection{Dust Continuum Measurements}

We measure the dust continuum flux densities of the DSFGs using the available ALMA continuum images and Herschel/SPIRE data. 
Fig.~\ref{fig:finding_chart} presents composite ALMA and JWST views of these sources. 
The DSFGs are covered by multiple ALMA bands from Band~3 to Band~8 (up to six bands), and the multi-band continuum images are shown in Appendix~\ref{app:cont_images}.
Flux measurements are performed following the methods adopted in previous studies 
(e.g., \cite{2017ApJ...835...98U}; \yearcite{2018PASJ...70...65U}; \yearcite{2026ApJ...997...79U}; \cite{2025A&A...699A.324H}). 
In brief, we primarily perform two-dimensional Gaussian fitting in the image plane using the \textsc{casa} task \texttt{imfit}. 
The resulting model and residual images are visually inspected for each source. 
For sources that exhibit internal structure leading to significant residuals, we instead measure flux densities within $2\sigma$ isophotal contours (\cite{2026ApJ...997...79U}). 
\textcolor{blue}{For non-detections, we report $3\sigma$ upper limits estimated
within a $3^{\prime\prime}$-diameter aperture. The aperture size
was motivated by curve-of-growth tests, which indicate that
this aperture typically encloses the majority of the line flux
for detected sources. The corresponding noise level was
estimated by scaling the image r.m.s.\ according to the aperture-to-beam area ratio.}

In addition to the ALMA data, we make use of Herschel/SPIRE observations at 250, 350, and 500\,$\mu$m in the SSA22 field (\cite{2016MNRAS.460.3861K}). 
These data provide photometric constraints at shorter wavelengths not covered by ALMA. 
As described in previous works (e.g., \cite{2020MNRAS.494.3828D}), Herschel flux densities (or upper limits) for individual ADF22 DSFGs are derived through a deblending analysis following \citet{2014MNRAS.438.1267S}, 
using the positions of ALMA sources, \textit{Spitzer}/MIPS 24\,$\mu$m sources, and 1.4\,GHz radio sources as priors (\cite{2014MNRAS.440.3462U}).
The quoted uncertainties and upper limits are estimated through Monte Carlo deblending and source injection simulations following \citet{2014MNRAS.438.1267S}, taking into account the instrumental noise and source confusion in the SPIRE maps.
We combine the ALMA and Herschel photometry to construct a comprehensive submillimeter–millimeter catalog covering observed wavelengths from 250\,$\mu$m to 3.26\,mm, corresponding to rest-frame $\sim60$--800\,$\mu$m at $z=3.09$.

\subsubsection{Methods}

We model the dust continuum emission using an optically thin modified blackbody (MBB) model,
implemented in the Bayesian SED-fitting code \textsc{mercurius}
(\cite{2022MNRAS.515.1751W}; \yearcite{2023MNRAS.523.3119W}).
In the SED fitting, we include absolute flux calibration uncertainties
(5\% for Bands~3--5 and 10\% for Bands~6--8), in addition to the statistical uncertainties derived from the image-plane measurements, following the ALMA Technical Handbook.
We assume optically thin dust emission and initially allow both the dust temperature ($T_{\rm dust}$)
and the dust emissivity index ($\beta$) to vary.
We adopt prior distributions consistent with those used in previous studies
(e.g., \cite{2023MNRAS.523.3119W,2024A&A...685A.138V}).
For $T_{\rm dust}$, we employ the default gamma prior in \textsc{mercurius} with a shape parameter $a = 1.5$,
truncated at the cosmic microwave background temperature ($T_{\rm CMB}$).
For $\beta$, 
we adopt a Gaussian prior of $\beta = 1.8 \pm 0.25$, following the default setting of \textsc{mercurius}. This choice is broadly consistent with recent observational studies of DSFGs (e.g., \cite{2024MNRAS.530.4887W}).
The effect of the cosmic microwave background (CMB) on the dust SED is explicitly included in the modeling (\cite{2015ApJ...806..110D}).
To convert the fitted SED normalization into dust mass, \textcolor{blue}{we adopts a dust mass absorption coefficient,
$\kappa_\nu = \kappa_0 (\nu / \nu_0)^{\beta}$,
with $\kappa_0 = 0.77$\,m$^2$\,kg$^{-1}$ at $\nu_0 = 850$\,GHz
(\cite{2000MNRAS.315..115D})}.
For sources with fewer than three continuum detections, we fix the emissivity index to $\beta = 2$,
which is representative of the best-fitting values obtained for sources with three or more continuum detections.

\subsubsection{Results}

The best-fit dust SEDs of the 18 ADF22 DSFGs at $z\simeq3.09$ are shown in Fig.~\ref{fig:sed}, and the derived physical parameters—far-infrared luminosity ($L_{\rm FIR}$; $42.5$–$122.5\,\mu$m), dust mass ($M_{\rm d}$), dust temperature ($T_{\rm dust}$), peak temperature ($T_{\rm peak}$; \textcolor{blue}{converted from the SED peak wavelength)}, and emissivity index ($\beta$)—are summarized in Table~\ref{tab:dust_stellar_properties}. 
Our $1.17\,\mathrm{mm}$ flux-limited sample spans approximately two orders of magnitude in far-infrared luminosity, $L_{\rm FIR}\sim10^{11}$--$10^{13}\,L_\odot$, with a median of $\log(L_{\rm FIR}/L_\odot)=11.56^{+0.32}_{-0.37}$.
The derived emissivity indices range from $\beta \simeq 1.8$–2.2, consistent with previous studies of DSFGs across a wide redshift range \textcolor{blue}{(e.g., \cite{
2018PASJ...70L...6K,2020ApJ...902...78R,2021ApJ...919...30D,2023MNRAS.523.3119W,2024MNRAS.530.4887W,2025MNRAS.540.1560B})}.

Fig.~\ref{fig:fir_td} shows the relation between $L_{\rm FIR}$ and $T_{\rm dust}$. 
We compare our sample with SPT-selected DSFGs at $z_{\rm spec}=3.1\pm0.5$ and AS2UDS-selected DSFGs at $z_{\rm phot}=3.1\pm0.5$, whose SEDs are constrained by ALMA and Herschel/SPIRE data \citep{2020ApJ...902...78R,2020MNRAS.494.3828D}. 
The combined data show a broad positive correlation between far-infrared luminosity and dust temperature, reflecting more efficient dust heating in more luminous systems.
The ADF22 DSFGs occupy a lower-luminosity regime ($L_{\rm FIR} \sim 10^{11}$--$10^{13}\,L_\odot$) compared to the SPT-selected sample. 
The median dust temperature is $T_{\rm dust}=25.0_{-3.2}^{+5.8}$\,K, significantly lower than that of the SPT DSFGs ($T_{\rm dust}=46.0_{-7.6}^{+14.3}$\,K). 
The AS2UDS-selected DSFGs have $T_{\rm dust}=27.2_{-2.3}^{+8.8}$\,K, broadly consistent with the ADF22 DSFGs, although their sample \textcolor{blue}{consists of galaxies with} higher $L_{\rm FIR}$.
These results suggest that the ADF22 sample probes a relatively faint and cool DSFG population at $z\sim3$ \textcolor{magenta}{, spanning both the LIRG and ULIRG regimes}, likely more representative of the bulk of star-forming galaxies, \textcolor{blue}{rather than the more luminous and potentially more burst-dominated systems preferentially selected in wide-area surveys}.

\subsection{Correlation with Galaxy Morphologies}

Among the 18 DSFGs in our sample, eight are detected in both the F444W and 870\,$\mu$m images at a spatial resolution of $0.15^{\prime\prime}$ (\cite{2026ApJ...997...79U}). 
This enables a direct comparison between the dust properties derived from the SED fitting and the structural parameters of the stellar and dust components. 
We adopt the effective radii reported in \citet{2026ApJ...997...79U} and stellar masses from \citet{2025A&A...699A.324H}. 
We calculate the stellar surface density ($\Sigma_\star$), dust surface density ($\Sigma_{\rm dust}$), and far-infrared surface brightness ($\Sigma_{\rm FIR}$) as 
$\Sigma_\star = M_\star/(2\pi R_{\rm e,JWST}^2)$, 
$\Sigma_{\rm dust} = M_{\rm dust}/(2\pi R_{\rm e,dust}^2)$, and 
$\Sigma_{\rm FIR} = L_{\rm FIR}/(2\pi R_{\rm e,dust}^2)$, 
\textcolor{blue}{where $R_{\rm e,JWST}$ and $R_{\rm e,dust}$ are the circularized effective radii measured from the F444W (or F560W/F770W) and 870\,$\mu$m images, respectively}. 

\textcolor{blue}{Throughout this analysis, we assume that the F444W emission traces the stellar mass distribution and that the 870\,$\mu$m emission traces the dust mass distribution on these spatial scales. 
While this assumption provides a useful framework for comparing the stellar and dust distributions, the F444W sizes may still be affected by dust attenuation, which could lead to an overestimation of the effective radii. 
If so, the intrinsic stellar surface densities would be even higher than estimated here. 
The morphology of the dust continuum emission may also be affected by spatial variations in dust temperature, such that the 870\,$\mu$m emission does not necessarily trace the dust mass distribution in a one-to-one manner. With these cautions in mind, we proceed to the following discussion.
}

The resulting spatially resolved view is shown in Fig.~\ref{fig:sigma_fir_td}. 
Fig.~\ref{fig:sigma_fir_td}a presents the relation between dust temperature ($T_{\rm dust}$) and far-infrared surface brightness ($\Sigma_{\rm FIR}$). 
In the optically thin limit, this relation is expected to follow $\Sigma_{\rm FIR} \propto \Sigma_{\rm dust} T_{\rm d}^{4+\beta}$. 
A model track with $\beta=2$, normalized to the median $\Sigma_{\rm dust}$ of the sample, is overlaid. 
\textcolor{blue}{The observed trend is broadly consistent with the expectation from optically thin modified blackbody models, in which higher dust temperatures correspond to higher FIR surface brightness at fixed dust surface density. At fixed $\Sigma_{\rm FIR}$, systems with larger $\Sigma_{\rm dust}$ are expected to exhibit lower $T_{\rm dust}$ under the optically thin assumption, consistent with the distribution seen in Fig.~\ref{fig:sigma_fir_td}a}.
\textcolor{blue}{In particular, some galaxies reach high $\Sigma_{\rm FIR}$ while maintaining moderate $T_{\rm dust}$, despite their compact and bright 870\,$\mu$m emission.
This suggests that large dust column densities act to moderate the local radiation field, thereby shaping the $\Sigma_{\rm FIR}$–$T_{\rm dust}$ relation.}

To further investigate this, we examine the relation between stellar and dust surface densities (Fig.~\ref{fig:sigma_fir_td}b). 
We find a positive correlation between $\Sigma_\star$ and $\Sigma_{\rm dust}$.
\textcolor{blue}{
We find a tentative positive trend between $\Sigma_\star$ and $\Sigma_{\rm dust}$, although the scatter is substantial and the current sample size is limited. 
This behavior is consistent with a scenario in which the concentration of stars and cold ISM increases in tandem, as gas is funneled toward the central regions during a compaction phase, triggering intense central star formation and the build-up of compact stellar structures.
}
The most stellar-mass–dense systems, ADF22.A4 and ADF22.A6, lie in the compact star-forming galaxy regime 
($\log \Sigma_\star \gtrsim 9.5\,{\rm M_\odot\,kpc^{-2}}$; \cite{2017ApJ...840...47B}), 
suggesting that they are undergoing a compaction phase. 
These galaxies also exhibit large dust surface densities, consistent with an enhanced concentration of the cold ISM.


%
Taken together, the correlations among $\Sigma_{\rm FIR}$, $T_{\rm dust}$, $\Sigma_\star$, and $\Sigma_{\rm dust}$ suggest that the spatial build-up of stellar mass and cold ISM are closely linked in DSFGs undergoing active star formation.
This picture implies that the dust distribution traces the broader assembly of baryonic mass rather than being confined to a purely central component.
%
%
While the dust emission is generally more compact than the stellar component ($R_{\rm e,dust} < R_{\rm e,JWST}$) (e.g., \cite{2025ApJ...978..165H,2026ApJ...997...79U}), this relative compactness does not necessarily imply that star formation is restricted to the central regions ($r\lesssim1$\,kpc). 
Instead, it reflects a higher central concentration of dust relative to stars, \textcolor{blue}{whereas the correlation in surface densities suggests that regions with high dust and gas concentrations are also the sites of active stellar mass assembly over galactic scales}.
In this sense, size-based comparisons alone provide a limited view of the underlying structure, and may be further affected by surface-brightness sensitivity, which preferentially traces the brightest components. 
The results for the ADF22 DSFGs therefore suggest that dust heating properties and structural evolution are closely linked through the coupled growth of baryonic components. 
This highlights the importance of sensitivity and spatial resolution in interpreting dust morphologies.
Mass surface densities provide a more direct measure of the physical conditions governing star formation than size alone.

\section{Molecular Gas Views}

\begin{figure*}
  \begin{center}
\includegraphics[width=15.5cm]
{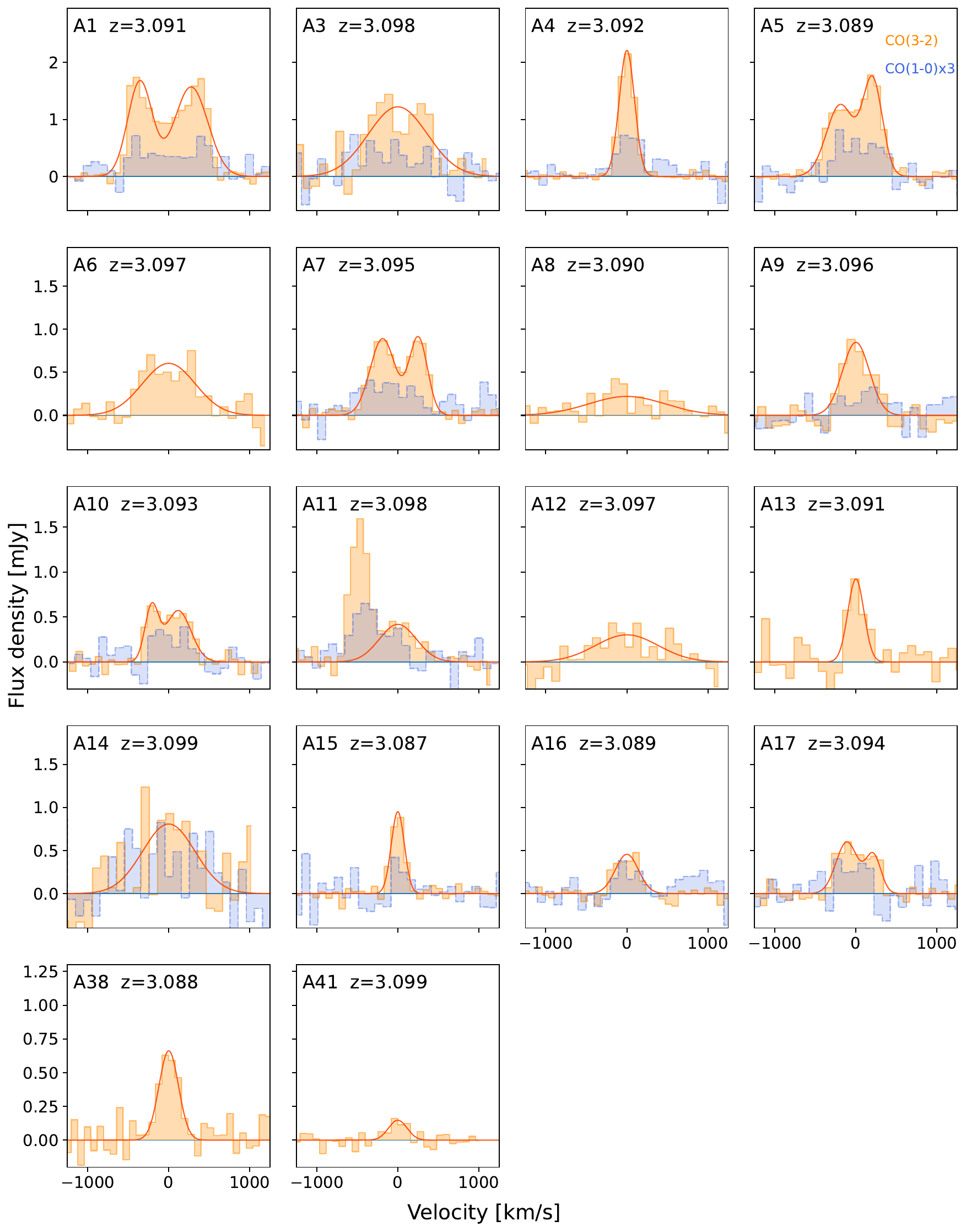}
  \end{center}
  \caption{
CO(3--2) and CO(1--0) spectra of the 18 DSFGs at $z\simeq3.09$ in the ADF22 field. 
The CO(3--2) spectra are shown in orange, extracted using apertures defined by the \texttt{imfit} positions and convolved source sizes, and the best-fit Gaussian models are overplotted with orange curves. 
If the CO(1--0) line is detected, the spectra are shown in blue, extracted using apertures corresponding to the synthesized beam size (for clarity, the flux density is multiplied by a factor of three). 
The velocity axis is defined relative to the systemic redshift derived from the CO(3--2) line. 
\textcolor{blue}{The bluer components in the A11 panel comes from A4 and hence not included in the fit.}
CO(3--2) emission is detected in all 18 galaxies, while CO(1--0) is detected in 12 DSFGs, providing a comprehensive view of the molecular gas reservoirs in DSFGs within the $z\simeq3.09$ proto-cluster core.
{Alt text: spectra.} 
}%
  \label{fig:spectra}
\end{figure*}

\subsection{Molecular line measurements}

\subsubsection{CO(3--2) detection}

We search for CO(3--2) emission using the ALMA Band~3 data cubes, including the ADF22 mosaic and the AzTEC1 and AzTEC14 deep fields (Fig.~\ref{fig:fov}). 
Among the 18 DSFGs at $z_{\rm spec}\simeq3.09$, \citet{2019Sci...366...97U} reported spectroscopic redshifts for 16 sources (A1 and A3--A17), based on CO(3--2) detections in the ADF22 mosaic and/or [O\,{\sc iii}]~$\lambda5008$ emission detected with Keck/MOSFIRE $K$-band spectroscopy (see also \cite{2015ApJ...815L...8U, 2017PASJ...69...45H, 2017ApJ...835...98U, 2018PASJ...70...65U}). 
\citet{2025A&A...699A.324H} identified two additional proto-cluster members (A38 and A41) based on CO(3--2) detections in the AzTEC1 deep field. 
In this work, we re-analyze three sets of Band~3 observations that cover the redshifted CO(3--2) emission at $z\approx3.09$.
For each source, we adopt the dataset with the highest sensitivity. When a source is covered by the deep fields, we use the corresponding data; otherwise, we use the ADF22 mosaic.

To measure the CO(3--2) flux, we first construct collapsed line maps using the \texttt{immoments} task in CASA over the velocity channels where line emission is detected. 
We then perform two-dimensional Gaussian fitting on the image plane using the \texttt{imfit} task, which provides the source coordinates, convolved and deconvolved sizes (FWHM), and integrated line flux. 
Spectra are extracted using apertures centered on the fitted positions, with sizes corresponding to the convolved source sizes. 
We fit the extracted spectra with Gaussian models to derive the systemic redshift and velocity widths. 
The number of Gaussian components (one or two) is determined using the Bayesian Information Criterion (BIC),
which selects the preferred model by balancing goodness of fit and model complexity.
CO(3--2) emission is detected in all 18 DSFGs. 
The derived parameters are summarized in Table~\ref{tab:co32co10}, and the extracted spectra and best-fit models are shown in Fig.~\ref{fig:spectra}.

\subsubsection{CO(1--0) measurements}

We search for CO(1--0) emission using the JVLA Ka-band data cubes, guided by the CO(3--2) measurements. 
Among the 18 DSFGs, 17 are covered by one of the three JVLA Ka-band datasets with a primary beam response of $>0.3$ at the source position. We construct collapsed images over the velocity range where line emission is detected or expected based on the CO(3--2) results.
Given the relatively coarse angular resolution and limited signal-to-noise ratio of the Ka-band data, we adopt different measurement strategies depending on source brightness. 
For the brightest sources (A1, A3, A4, A5, A7, and A9), we measure the flux using the \texttt{imfit} task. 
For the remaining sources, we extract spectra using apertures centered on the CO(3--2) positions with sizes corresponding to the synthesized beam, and measure the line flux by fitting the spectra while fixing the velocity center and width to those derived from CO(3--2).
For non-detections, we adopt $5\sigma$ point-source upper limits based on the collapsed images. 
As a result, CO(1--0) emission is detected in 12 of the 17 sources observed with the JVLA Ka-band. 
The spectra are shown in Fig.~\ref{fig:spectra}, and the derived parameters are summarized in Table~\ref{tab:co32co10}.

\subsubsection{CO(8--7), CO(9--8), and CO(12--11) measurements}

\textcolor{blue}{Among the ADF22 DSFGs, four sources (ADF22.A1, A4, A6, and A7) were targeted for CO(8--7) and CO(9--8) observations. 
As shown in Fig.~\ref{fig:highjspec}, both lines are detected in all four DSFGs. 
We also searched for the CO(12--11) emission line using the Band~7 data cubes, which were originally obtained for the 870\,$\mu$m continuum observations, and identified CO(12--11) emission from ADF22.A4.
For the detected cases, we constructed collapsed line maps using the \texttt{immoments} task over the same velocity ranges adopted for the CO(3--2) analysis, and measured the line fluxes with \texttt{imfit}. 
For the non-detections of CO(12--11), we report $3\sigma$ upper limits measured within a $3^{\prime\prime}$ aperture. 
The resulting line properties are summarized in Table~\ref{tab:highj_co}}. 

\subsection{Brightness temperature ratio $r_{31}$}

We calculate CO line luminosities from the integrated line fluxes following \citet{2005ARA&A..43..677S}:
\begin{equation}
L'_{\rm CO} = 3.25\times10^7\, I_{\rm CO}\, D_{\rm L}^2\, (1+z)^{-3}\, \nu_{\rm obs}^{-2}
\end{equation}
where $I_{\rm CO}$ is the integrated line flux in Jy\,km\,s$^{-1}$, $\nu_{\rm obs}$ is the observed frequency in GHz, and $D_{\rm L}$ is the luminosity distance in Mpc. 
We define the brightness temperature ratio as $r_{31}=L'_{\rm CO(3-2)}/L'_{\rm CO(1-0)}$, which traces the average excitation state of the molecular gas in the low-$J$ regime. 
The derived values are summarized in Table~\ref{tab:co32co10}.

\textcolor{blue}{For the 12 ADF22 DSFGs with direct CO(1--0) detections, we obtain a median line ratio of 
$r_{31}=0.66^{+0.05}_{-0.04}$. 
Including the five sources with CO(1--0)-based upper limits by uniformly sampling the allowed range below each limit yields a slightly lower median value of 
$r_{31}=0.59^{+0.05}_{-0.05}$. 
We therefore adopt the detection-based median as our fiducial value, while noting that the inclusion of upper limits does not significantly alter the representative excitation correction.}
To estimate a representative value, we perform Monte Carlo sampling that accounts for the measurement uncertainties of individual galaxies. 
For each realization, $r_{31}$ values are drawn from asymmetric Gaussian distributions defined by the measured values and their uncertainties, and the sample median is computed. 
The final representative value and its uncertainty are taken from the 50th, 16th, and 84th percentiles of the resulting distribution of median values.
This value is consistent with $r_{31}=0.63\pm0.12$ derived from 39 DSFGs compiled by \citet{2021MNRAS.501.3926B} (see also \cite{2011MNRAS.412.1913I,2013MNRAS.429.3047B,2016ApJ...827...18S,2020ApJ...896L..21R,2023ApJ...945..128F}), suggesting that the average excitation properties of ADF22 DSFGs are similar to those of DSFGs in general fields.

\textcolor{blue}{Among the ADF22 DSFGs, six sources (ADF22.A1, A4, A6, A7, A9, and A12) are known to host X-ray AGNs \citep{2009ApJ...691..687L,2010ApJ...724.1270T,2015ApJ...815L...8U,2019Sci...366...97U}. 
To investigate the possible influence of AGN activity on the molecular gas excitation, we divide the sample into X-ray AGN and non-AGN subsamples and compare their $r_{31}$ values. 
Using only sources with direct CO(1--0) detections, we obtain median values of 
$r_{31}=0.65_{-0.04}^{+0.04}$ and $0.66_{-0.08}^{+0.10}$ 
for the AGN and non-AGN subsamples, respectively. 
Including the sources with CO(1--0) upper limits yields median values of 
$r_{31}=0.61_{-0.04}^{+0.04}$ and $0.55_{-0.10}^{+0.09}$, respectively. 
Within the current uncertainties, no significant difference is found between the two subsamples. 
This result suggests that the presence of X-ray AGNs does not strongly affect the global low-to-mid-$J$ CO excitation of the molecular gas reservoirs in these DSFGs, consistent with previous studies \citep{2016ApJ...827...18S}.}

\citet{2020ApJ...896L..21R} reported that five galaxies at $z\sim2$--3 selected via CO(3--2) detections in the ASPECS survey have $r_{31}=0.84\pm0.26$, and suggested that CO(3--2)-selected galaxies tend to exhibit higher excitation than CO(1--0)-selected systems. 
If we consider only the five DSFGs with the highest CO(3--2) luminosities in our sample, we obtain $r_{31}=0.72_{-0.07}^{+0.14}$, which is slightly higher than the median value for the full ADF22 sample 
\textcolor{blue}{, suggesting a possible selection bias toward more highly excited systems in CO(3--2)-selected samples, although larger samples are needed to verify this trend}.

\subsection{$\alpha_{\rm CO}$ and molecular gas mass}

\begin{figure}
  \begin{center}
\includegraphics[width=8.5cm]
{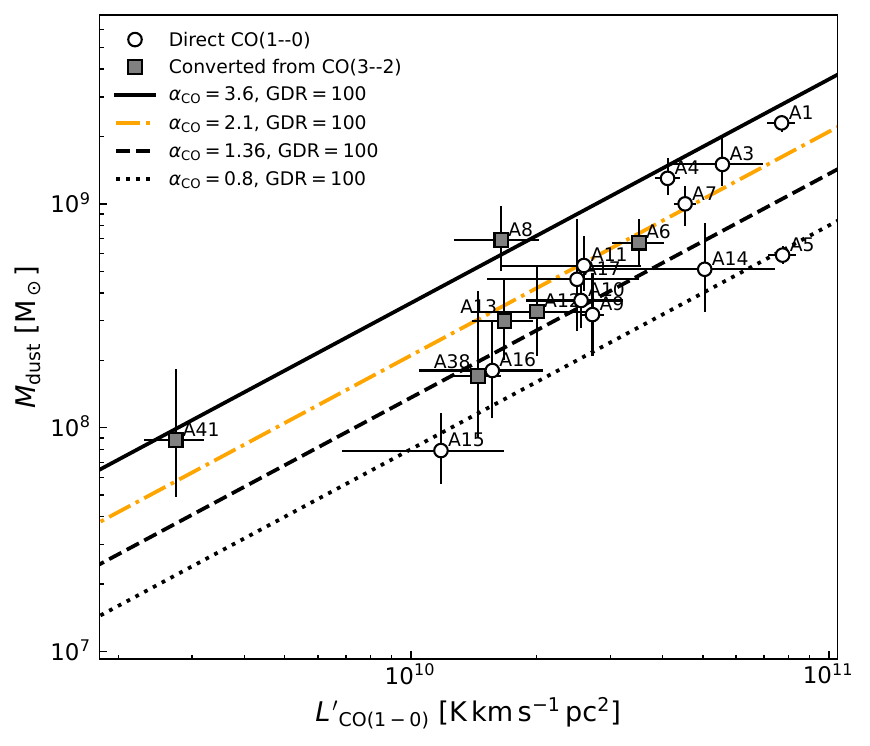}
  \end{center}
  \caption{
Relation between CO(1--0) luminosity and dust mass for the ADF22 DSFGs.
Open circles denote galaxies with direct CO(1--0) detections, while filled squares indicate sources for which $L'_{\rm CO(1-0)}$ is inferred from CO(3--2) using the median excitation ratio of the sample.
The black solid line shows the relation expected for $\alpha_{\rm CO}=3.6$ and $\mathrm{GDR}=100$.
For comparison, the black dotted and dashed lines show the relations for $\alpha_{\rm CO}=0.8$ and $\alpha_{\rm CO}=1.36$, respectively.
The orange dash--dotted line represents the relation implied by the median $M_{\rm dust}/L'_{\rm CO(1-0)}$ ratio measured for the ADF22 DSFGs ($\alpha_{\rm CO}=2.1$ for $\mathrm{GDR}=100$).
\textcolor{blue}{Most sources lie within the region encompassed by $\alpha_{\rm CO}=0.8$--3.6, although substantial source-to-source variations are also present.}
{Alt text: diagram of CO luminosity and dust mass.} 
}%
  \label{fig:LpcoMd}
\end{figure}

\begin{figure*}
  \begin{center}
\includegraphics[width=17cm]
{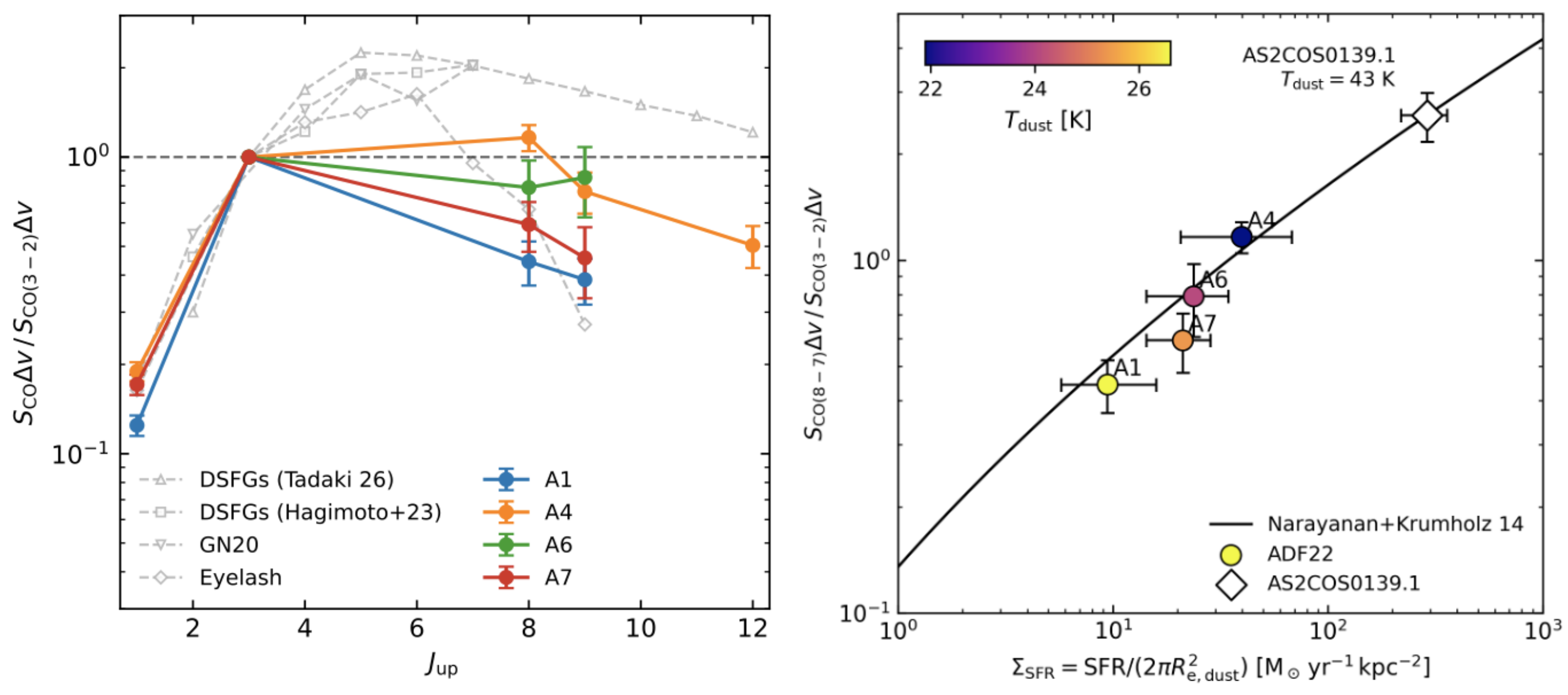}
  \end{center}
  \caption{
\textcolor{blue}{(left) CO SLEDs of the four ADF22 DSFGs with high-$J$ CO detections, normalized to the CO(3--2) transition. For comparison, the CO SLEDs of a compilation of DSFGs \citep{2023MNRAS.521.5508H,2026arXiv260223521T} and two individual DSFGs \citep{2011MNRAS.410.1687D,2026ApJ...996...19B} are shown with the same normalization. The high-$J$ CO lines of the ADF22 DSFGs generally exhibit relatively low excitation compared to other DSFGs, although the high-$J$ excitation regime remains relatively poorly constrained. 
(right) Relation between $\Sigma_{\rm SFR}$ and the flux ratio $S_{\rm CO(8-7)}\Delta v / S_{\rm CO(3-2)}\Delta v$ for the four ADF22 DSFGs, compared with another DSFG (\cite{2025MNRAS.536.1149T}) and the theoretical model of \citet{2014MNRAS.442.1411N}. The ADF22 DSFGs are color-coded by $T_{\rm dust}$, while AS2COS0139.1 is shown separately with its dust temperature indicated. The observed values generally follow the model track, suggesting that the molecular gas excitation in the ADF22 DSFGs is primarily regulated by intense star formation.}
{Alt text: CO SLEDs.} 
}%
  \label{fig:co87}
\end{figure*}

We estimate the total molecular gas mass, $M_{\rm gas}$, from the CO luminosity as
\begin{equation}
M_{\rm gas} = \alpha_{\rm CO} L'_{\rm CO(1-0)},
\end{equation}
where $\alpha_{\rm CO}$ is the CO-to-molecular-gas conversion factor, including the contribution from helium. 
For sources without CO(1--0) detections, we estimate $L'_{\rm CO(1-0)}$ from CO(3--2) using the median $r_{31}$ value of the sample.

\textcolor{blue}{The choice of $\alpha_{\rm CO}$ remains a major source of uncertainty in deriving molecular gas masses, as it depends on physical conditions such as metallicity and gas distribution.}
For example, some studies adopt $\alpha_{\rm CO}=3.6$ for massive main-sequence galaxies (e.g., \cite{2010ApJ...713..686D,2020ApJ...896L..21R,2019ApJ...882..140B}), whereas others adopt $\alpha_{\rm CO}=1.36$ for massive galaxies selected as DSFGs (e.g., \cite{2021MNRAS.501.3926B,2023ApJ...945..128F}), \textcolor{blue}{corresponding to $\alpha_{\rm CO}=1$ multiplied by a factor of 1.36 to include helium}.

The fact that DSFGs are intensely star-forming systems yet often overlap with the main-sequence population (e.g., \cite{2015ApJ...806..110D,2025A&A...699A.324H}), together with the difficulty of measuring gas-phase metallicities for such systems, further complicates the interpretation.
This highlights the importance of empirically constraining $\alpha_{\rm CO}$ for each sample.
In this section, we assess the appropriate $\alpha_{\rm CO}$ for the ADF22 DSFGs by comparing CO luminosities with dust masses derived from infrared SED fitting using multi-band ALMA photometry.

Fig.~\ref{fig:LpcoMd} compares the CO(1--0) luminosities with the dust masses. 
The ADF22 DSFGs exhibit a correlation between these quantities \textcolor{blue}{although substantial source-to-source variations are also present}.
The median ratio is 
\textcolor{blue}{$M_{\rm dust}/L'_{\rm CO(1-0)} = 0.021^{+0.003}_{-0.003}$}. 
Assuming a characteristic gas-to-dust ratio of $\mathrm{GDR}=100$ (\cite{2014MNRAS.438.1267S,2016ApJ...820...83S}), this implies 
\textcolor{blue}{$\alpha_{\rm CO} = 2.1^{+0.3}_{-0.3}$}. 
This value is between the two cases; the main-sequence conversion factor ($\alpha_{\rm CO}\approx3.6$; e.g., \cite{2010ApJ...713..686D}) and the ULIRG-like value ($\alpha_{\rm CO}\approx0.8$; \cite{1998ApJ...507..615D}). 
\textcolor{blue}{We note that this comparison does not uniquely constrain $\alpha_{\rm CO}$ given the various uncertainties involved, including the gas-to-dust ratio and dust mass absorption coefficient}. Nevertheless, the comparison suggests that the molecular gas in the ADF22 DSFGs is not necessarily in the extreme starburst regime typically associated with merger-driven systems. 
The ADF22 DSFGs lie on the $M_\star$--SFR relation of typical star-forming galaxies (\cite{2025A&A...699A.324H}) and exhibit predominantly disk-like stellar morphologies (\cite{2026ApJ...997...79U}). Motivated by these properties, we adopt $\alpha_{\rm CO}=3.6$ to estimate the molecular gas masses. 
The derived values are summarized in Table~\ref{tab:molgas}. 
The ADF22 DSFGs span more than one order of magnitude in molecular gas mass, ranging from 
$M_{\rm gas}=1\times10^{10}$ to $2.8\times10^{11}\,M_\odot$.

\subsection{High-$J$ CO Excitation}

\textcolor{blue}{
For the four DSFGs (ADF22.A1, A4, A6, and A7), we construct CO SLEDs spanning CO(1--0), CO(3--2), CO(8--7), CO(9--8), and CO(12--11), as shown in the left panel of Fig.~\ref{fig:co87}. 
The SLEDs are normalized by the CO(3--2) flux ($S_{\rm CO(3-2)}\Delta v$). 
For comparison, the CO SLEDs of a compilation of DSFGs \citep{2023MNRAS.521.5508H} \textcolor{magenta}{, a sample biased toward high-excitation DSFGs \citep{2026arXiv260223521T}}, and two individual DSFGs \citep{2011MNRAS.410.1687D,2026ApJ...996...19B} are shown with the same normalization. 
Although the lack of mid-$J$ CO measurements leaves significant uncertainties in the detailed shape of the SLEDs and prevents detailed excitation modeling, the relation between the high-$J$ CO lines and CO(3--2) suggests that the ADF22 DSFGs generally exhibit relatively low excitation compared to other DSFGs.
}

\textcolor{blue}{
\citet{2014MNRAS.442.1411N} showed that CO excitation correlates with the star-formation rate surface density, as higher $\Sigma_{\rm SFR}$ leads to both higher gas densities and stronger heating in the molecular ISM, which together enhance the population of high-$J$ CO levels. 
In the right panel of Fig.~\ref{fig:co87}, we show the relation between $\Sigma_{\rm SFR}$ and the CO(8--7)/CO(3--2) flux ratio ($S_{\rm CO(8-7)}\Delta v / S_{\rm CO(3-2)}\Delta v$), with the marker color indicating $T_{\rm dust}$. 
Here we calculate the star-formation rate surface density as 
$\Sigma_{\rm SFR}={\rm SFR}/(2\pi R_{{\rm e},{\rm dust}}^2)$, 
where the SFR is derived from the SED fitting (\cite{2025A&A...699A.324H}) and $R_{{\rm e},{\rm dust}}$ is the circularized effective radius measured from the ALMA 870\,$\mu$m continuum emission. 
We also include the DSFG AS2COS0139.1 at $z=3.292$ from \citet{2025MNRAS.536.1149T}\textcolor{magenta}{, which enables us to expand the $\Sigma_{\rm SFR}$ coverage}.  
Since CO(3--2) measurements are unavailable for this source, we estimate the CO(3--2) flux from the observed CO(4--3) flux using the Eyelash template \citep{2011MNRAS.410.1687D}. 
For the theoretical prediction, we adopt the unresolved model relation of \citet{2014MNRAS.442.1411N}. 
}

The observed $\Sigma_{\rm SFR}$--$S_{\rm CO(8-7)}\Delta v/S_{\rm CO(3-2)}\Delta v$ relation is broadly consistent with the theoretical prediction of \citet{2014MNRAS.442.1411N} for the samples. 
In contrast, 
the current data do not show compelling evidence for a correlation between the CO(8--7)/CO(3--2) ratio and $T_{\rm dust}$, given the limited dynamic range, sample size, and uncertainties in $T_{\rm dust}$.
These results suggest that the molecular gas excitation in the ADF22 DSFGs is primarily regulated by intense star formation, and that the relatively low excitation of the ADF22 DSFGs is linked to their relatively low $\Sigma_{\rm SFR}$ values arising from their relatively extended dust emission. 
The absence of compelling evidence for a tight correlation between the CO(8--7)/CO(3--2) ratio and the galaxy-averaged dust temperature may indicate
that the excitation of high-$J$ CO lines is more directly linked to the local density and heating conditions of compact star-forming regions than to the global dust temperature of the galaxy.

\textcolor{blue}{
We note that all four ADF22 DSFGs host X-ray AGNs, raising the possibility that AGN-heated CO(8--7) components associated with X-ray dominated regions and/or shocks may contribute to the observed emission. 
While we do not exclude this possibility, the relatively low excitation compared to other DSFGs and the consistency with the star-formation-driven model of \citet{2014MNRAS.442.1411N} suggest that AGN heating does not dominate the overall CO SLEDs, even at these high-$J$ transitions. 
Meanwhile, the two sources exhibiting the largest CO(8--7)/CO(3--2) ratios also have the highest stellar surface densities. 
Taken together, these results may support a scenario in which centrally concentrated gas builds up a dense molecular phase, boosting the excitation of high-$J$ CO lines while simultaneously accompanying the build-up of compact stellar structures.
}

\section{Discussion}

\subsection{Star-formation and baryon cycle in proto-cluster DSFGs}

\begin{figure*}
  \begin{center}
\includegraphics[width=17cm]
{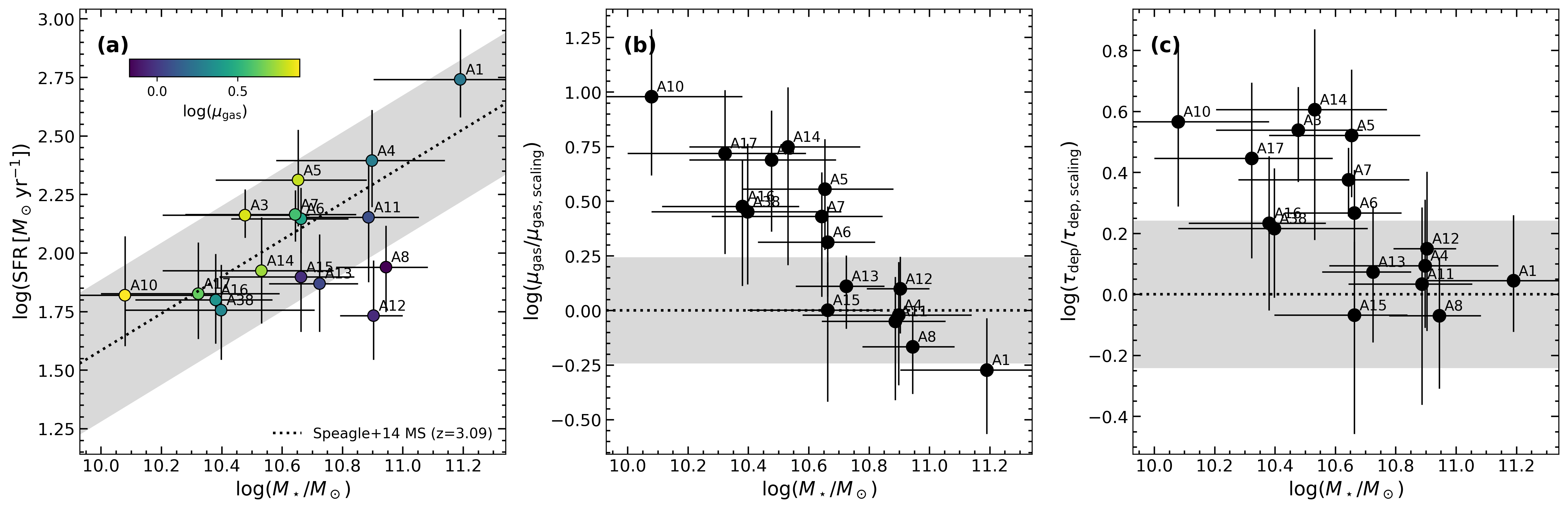}
  \end{center}
  \caption{
(a) Stellar mass–star formation rate relation for the ADF22 DSFGs at $z\simeq3.09$ (\cite{2025A&A...699A.324H}). The dashed line shows the star-forming main sequence from \citet{2014ApJS..214...15S} evaluated at ($z=3.09$), and the gray shaded region indicates its intrinsic scatter (0.3 dex). Colors represent the molecular gas fraction, ($\mu_{\rm gas}=M_{\rm mol}/M_\star$). Most sources lie within the main-sequence scatter, consistent with previous studies of DSFGs in this proto-cluster environment.
(b) Residual gas fraction relative to the field scaling relation of \citet{2018ApJ...853..179T}, shown as ($\log(\mu_{\rm gas}/\mu_{\rm gas,scaling}))$ as a function of stellar mass. The dashed horizontal line indicates the field scaling relation, and the gray band shows its intrinsic scatter. Massive systems broadly follow the field relation, whereas lower-mass galaxies exhibit systematically higher gas fractions than expected.
(c) Residual depletion time relative to the \citet{2018ApJ...853..179T} scaling relation, ($\log(\tau_{\rm dep}/\tau_{\rm dep,scaling}))$, where ($\tau_{\rm dep}=M_{\rm mol}/{\rm SFR}$). The gray band again represents the intrinsic scatter of the field relation. Similar to panel (b), lower-mass galaxies tend to show longer depletion times than predicted by the field scaling relations, indicating substantial molecular gas reservoirs relative to their current star-formation activity.
Taken together, these trends suggest that while massive DSFGs in ADF22 broadly follow the star-formation scaling relations observed in field galaxies, lower-mass systems exhibit enhanced gas fractions and longer depletion times, possibly reflecting ongoing gas accumulation in the proto-cluster environment.
{Alt text: Relations of molecular gas mass.} 
}%
  \label{fig:scaling}
\end{figure*}


Fig.~\ref{fig:scaling}a shows the location of the ADF22 DSFGs relative to the star-forming main sequence \citep{2014ApJS..214...15S}, with colors indicating the molecular gas fraction ($\mu_\mathrm{gas}=M_\mathrm{gas}/M_\star$). 
Fig.~\ref{fig:scaling}b and c compare the molecular gas fraction and depletion time with the field scaling relations of \citet{2018ApJ...853..179T}, \textcolor{blue}{which describe the typical gas properties of star-forming field galaxies as functions of stellar mass, redshift, and star-formation activity}.
\textcolor{blue}{Panels (b) and (c) provide complementary views: the gas fraction traces the accumulated molecular reservoir relative to the stellar mass, whereas the depletion time measures this reservoir relative to the current star-formation activity.}
We exclude ADF22.A9 and ADF22.A41 from the following analysis, as A9 is a QSO and A41 shows a potential spatial offset between the dust and stellar emission, making it unsuitable for a direct comparison.
Most ADF22 DSFGs lie within the intrinsic scatter (0.3~dex) of the main sequence at $z\simeq3.09$ \citep{2025A&A...699A.324H}.

\textcolor{blue}{We use the field scaling relations as a reference benchmark to investigate the behavior of galaxies in proto-cluster environments. 
As shown in Fig.~\ref{fig:scaling}b,c, a clear stellar-mass dependence emerges in both the gas fraction and depletion time. 
The excesses in gas fraction and depletion time relative to the field scaling relations are most prominent in the lower-mass regime ($\log (M_\star/M_\odot)\sim10.0$--10.6), whereas the massive systems ($\log (M_\star/M_\odot)\sim10.6$--11.2) broadly follow the field relations. 
Although the absolute normalization depends on the adopted $\alpha_{\rm CO}$, the relative trend with respect to the field scaling relations remains unchanged, with lower-mass systems exhibiting systematically larger excesses in gas fraction and depletion time than the massive galaxies.}

This trend is broadly consistent with the results of \citet{2019PASJ...71...40T}, who reported that protocluster galaxies at $z\sim2$--2.5 show larger gas fractions and longer depletion times than field scaling relations at $10.5 < \log(M_\star/M_\odot) < 11.0$, while more massive galaxies with $\log(M_\star/M_\odot) > 11.0$ are consistent with or slightly below the relations. 
They interpreted this behavior as evidence for a mass-dependent environmental effect, in which gas accretion along cosmic filaments is enhanced in less massive galaxies but suppressed in the most massive systems. 
A similar trend has been reported for another proto-cluster at $z\sim2.5$ \citep{2025A&A...698A.312G}. 
In contrast, \citet{2025A&A...701A.234Z} reported elevated gas fractions and depletion times even in massive galaxies with $M_\star \gtrsim 8\times10^{10}\,M_\odot$, indicating diversity in the gas properties of proto-cluster galaxies. 
\textcolor{blue}{They interpreted these gas-rich massive systems as evidence that substantial molecular gas reservoirs can sustain persistent star formation and early stellar mass assembly in proto-cluster cores}.

We note that these studies (e.g., \cite{2017ApJ...842...55L,2019PASJ...71...40T,2025A&A...701A.234Z}) are based primarily on CO(3--2) or CO(4--3) observations without constraints from lower-$J$ transitions, and adopt relatively large $\alpha_{\rm CO}$ values based on metallicity-dependent prescriptions inferred from stellar mass via the mass--metallicity relation. 
These assumptions introduce systematic uncertainties in molecular gas masses, which may contribute to the reported differences.
In particular, the conversion from mid- to high-$J$ CO lines to CO(1--0), together with the adopted $\alpha_{\rm CO}$, can significantly affect the inferred gas properties. 
This highlights the importance of comprehensive surveys that include multiple CO transitions, especially CO(1--0), combined with a consistent treatment of $\alpha_{\rm CO}$.

The ADF22 results can be interpreted within a simple picture in which lower-mass galaxies continue to be replenished by filamentary gas inflows, while the most massive systems have already evolved toward a more regulated state.
In addition, a comparable level of gas replenishment naturally produces a larger increase in gas fraction in lower-mass galaxies because of their smaller stellar masses.
%
In this scenario, the elevated gas fractions and longer depletion times of the lower-mass galaxies reflect a gas-rich growth phase, whereas the massive systems broadly follow the field scaling relations because a larger fraction of their molecular gas has already been consumed by star formation. 
At the same time, our data do not necessarily require suppressed gas supply in the massive systems. 
The massive DSFGs in ADF22 remain actively star-forming and gas-rich despite broadly following the field scaling relations, suggesting that their molecular gas reservoirs may already be regulated by a baryon cycle similar to that in field galaxies. 
In this interpretation, gas consumption by star formation may be approximately balanced by continued gas replenishment, maintaining a near-equilibrium state. 
A notable example is ADF22.A1, the most massive galaxy in the field, which remains gas-rich and lies on the main sequence. 
Taken together, the ADF22 results suggest that the stellar-mass dependence of gas fraction and depletion time in proto-cluster galaxies may reflect differences in the balance between gas supply and consumption, rather than a simple suppression of gas accretion in the most massive systems.

We note that our sample is limited to sources detected at $1.1\,\mathrm{mm}$ and may therefore preferentially select gas-rich galaxies. 
As a result, the present analysis does not provide a complete census of gas properties in the proto-cluster. 
Nevertheless, the observed stellar-mass dependence suggests that the environment influences the \textcolor{blue}{molecular gas content} of proto-cluster galaxies, particularly in lower-mass systems.
In the next section, we further explore this interpretation by examining gas surface densities and related structural properties.


\subsection{Evolution and transformation of massive galaxies along filaments}

\begin{figure*}
  \begin{center}
\includegraphics[width=15.5cm]
{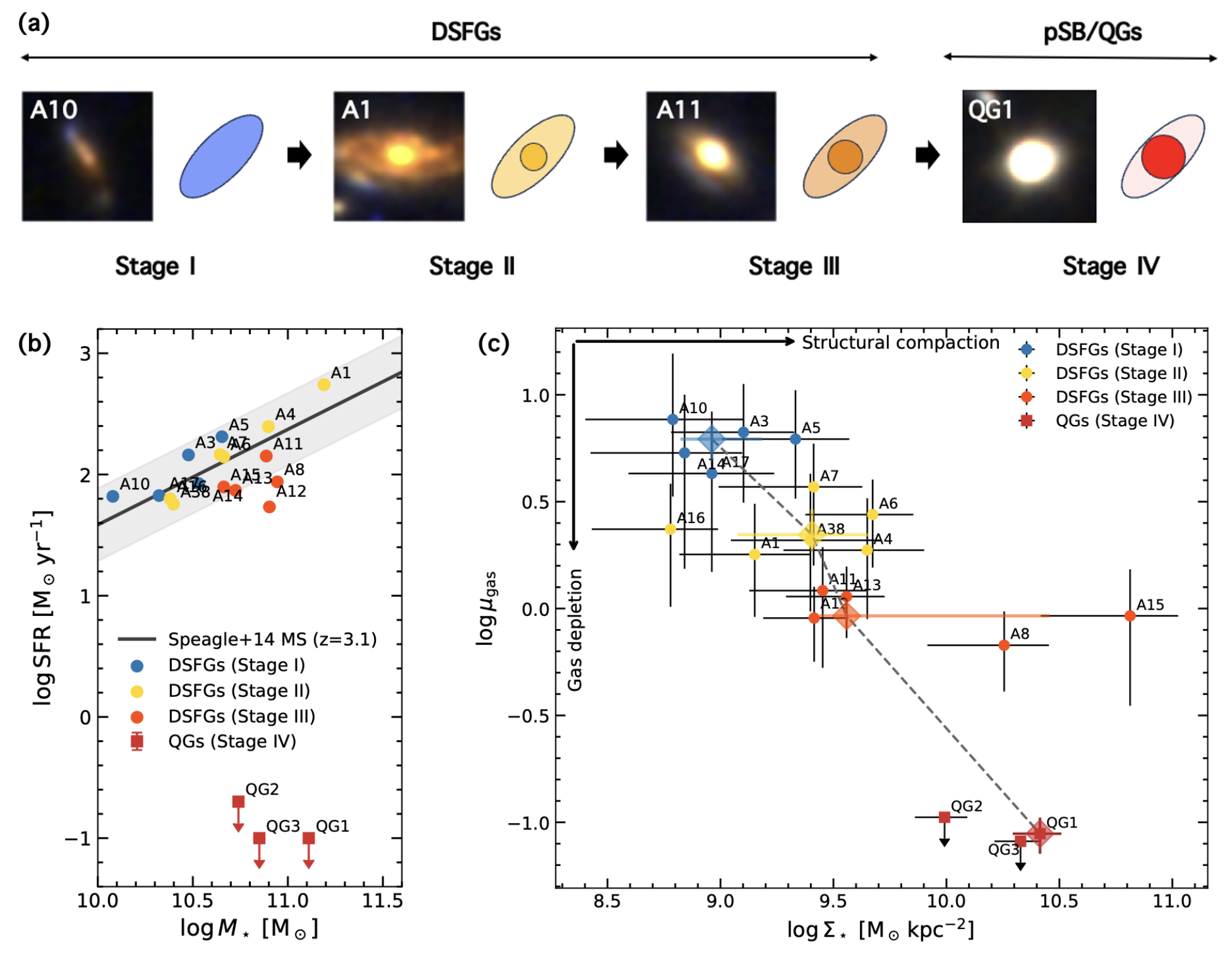}
  \end{center}
  \caption{
(a) A schematic picture of massive galaxy evolution in ADF22. 
An example NIRCam color image (F200W, F356W, F444W) is also shown. 
In Stage~I, galaxies are in the least evolved phase, with overabundant molecular gas fueled by the cosmic web. 
In Stage~II, galaxies experience peak star formation, where gas accretion from the cosmic web sustains the activity. 
In Stage~III, gas depletion proceeds while stellar mass becomes more centrally concentrated. 
In Stage~IV, galaxies reach the quiescent phase (QGs), in which molecular gas is largely depleted. 
(b) Stellar mass ($M_\star$)–SFR diagram showing the locations of galaxies in the four evolutionary stages, including QGs (see also Fig.~\ref{fig:scaling}a).
(c) Stellar mass surface density ($\Sigma_\star$)–gas mass fraction ($\mu_\mathrm{gas}$) diagram showing the locations of galaxies in the four evolutionary stages. 
The stage medians \textcolor{blue}{(large diamonds)} reveal a strong anti-correlation between gas fraction and stellar mass surface density, tracing a systematic transition from gas-rich DSFGs to compact quiescent systems. 
The arrows illustrate two orthogonal evolutionary trends: structural compaction (increasing $\Sigma_\star$) and gas depletion (decreasing $\mu_\mathrm{gas}$).
This diagram provides a unified view of gas consumption and structural evolution along the proposed sequence.
{Alt text: Stages and evolution.} 
}%
  \label{fig:evolution}
\end{figure*}

Using the comprehensive dataset covering the dust, gas, and stellar components, together with spatial information, we explore how massive galaxies form and evolve within the cosmic web at $z\approx3$.
%
Fig.~\ref{fig:evolution}a presents a schematic picture in which the DSFGs investigated in this work and the quiescent galaxies (QGs; \cite{2025ApJ...985L...8U}), all associated with Ly$\alpha$-traced cosmic web filaments, represent different stages of massive galaxy formation (Stage~I to Stage~IV). 
This classification is motivated by several observational clues (Fig.~\ref{fig:evolution}b,c). Although the ALMA-identified DSFGs lie close to the star-forming main sequence, they exhibit a wide range of stellar masses and gas fractions (Fig.~\ref{fig:scaling}), suggesting that different evolutionary stages are included. As suggested by \citet{2026ApJ...997...79U}, galaxies in the DSFG phase are also expected to undergo structural transformation, increasing their stellar mass concentration and bulge-to-total ratio. 

Less massive systems tend to be gas-rich and have a large gas mass fraction (Stage~I). As stellar mass builds up, galaxies increase their star-formation activity along the main-sequence track and reach a peak phase in which they show moderate gas mass fraction and stellar surface density (Stage~II). Subsequently, they show reduced gas fractions and further enhanced stellar surface density, suggesting a transition toward quiescence (Stage~III). Finally, passive populations with no evidence of ongoing star formation in any tracer represent the most evolved phase in which gas mass fraction is suppressed and stellar surface density is the most enhanced (Stage~IV). Thus, based on their location relative to the main sequence and their gas fractions, we introduce a four-stage evolutionary sequence from the least to the most mature systems.

To combine the gas consumption perspective with structural transformation, we construct a diagram of stellar surface density ($\Sigma_\star$) and gas fraction ($\mu_\mathrm{gas}$) in Fig.~\ref{fig:evolution}c. In this diagram, DSFGs and QGs form a sequence from the upper left to the lower right, reflecting decreasing gas fraction (downward) and increasing structural concentration (rightward). A natural interpretation is that we are witnessing an evolutionary sequence in which gas consumption and structural growth proceed in tandem.

The overall slope of the median values across the stages is close to unity, implying $\mu_\mathrm{gas} \propto \Sigma_\star^{-1}$ in a global sense. 
If the size ratio between the stellar component and molecular gas reservoir is approximately constant, this would suggest that the gas surface density remains roughly constant. However, this represents only a first-order approximation, and the actual evolutionary pathways are likely more complex. For example, Fig.~\ref{fig:sigma_fir_td}b indicates that the ISM surface density can evolve together with the stellar surface density. In contrast, the gas surface density may decrease again as gas depletion proceeds. The evolutionary track is therefore likely to involve a combination of horizontal and vertical movements in this diagram. Nevertheless, the overall trend supports a scenario in which galaxies evolve toward compact QGs through concurrent gas depletion and structural transformation.

\textcolor{blue}{Taken together, the observational properties of the ADF22 DSFGs—including their locations on the star-forming main sequence \citep{2025A&A...699A.324H}, relatively low dust temperatures, moderate CO excitation, and the preference for main-sequence-like $\alpha_{\rm CO}$ values—suggest that the DSFG overabundance in ADF22 is associated with a relatively regulated mode of star formation sustained by substantial molecular gas reservoirs. Sustained gas supply may contribute to mass assembly in tandem with mergers and interactions within the proto-cluster environment.}

Among the 18 DSFGs, six (A1, A4, A6, A7, A9, and A12) are identified as X-ray AGNs (\cite{2015ApJ...815L...8U,2019Sci...366...97U}; see also \cite{2009MNRAS.400..299L,2010ApJ...724.1270T,2021ApJ...919...51M,2023ApJ...951...15M}), indicating that black hole growth is also active in these systems. 
Five of these AGNs are classified as Stage~II, suggesting that active black hole growth broadly coincides with the phase of intense star formation (\cite{2009ApJ...696..891H,2020ApJ...892...17D}). 
In particular, A4 and A6 exhibit the highest stellar and dust surface densities, implying that gas inflow toward the central regions may be enhanced during the compaction phase. 
\textcolor{blue}{Among the three QGs, QG2 is also identified as an X-ray AGN (\cite{2009MNRAS.400..299L,2022ApJ...935...89K}), suggesting that AGN activity may persist even after quenching. 
We note that in some cases the measured stellar surface densities of DSFGs and QGs may be affected by AGN emission (for example, extremely compact sources like ADF22.A15), although the overall SEDs are likely dominated by stellar emission in most cases (e.g., \cite{2025PASJ...77..432U,2025A&A...699A.324H}).}

To examine the relation between evolutionary stage and location within the cosmic web, we calculate the line-of-sight velocity offsets and projected separations for two galaxy groups expected to reside in massive halos ($M_{\rm h}\gtrsim10^{13}\,M_\odot$; \cite{2016MNRAS.455.3333K,2025PASJ...77..432U}): AzTEC14 (A4, A10, A11, A16, A17, QG1–3) and AzTEC1 (A1, A5, A7, A15) in ADF22 \textcolor{blue}{(Appendix~\ref{Asec:mhalo} for details)}.
We find that the evolutionary stage correlates with position within the halo. 
In the AzTEC14 group, QGs are preferentially located near the center, while Stage~I galaxies reside in the outskirts. This trend is consistent with observations of a $z=2.5$ cluster core \citep{2018ApJ...867L..29W}. The relatively faint Ly$\alpha$ emission around AzTEC14 may indicate the emergence of a hot proto-ICM, which suppresses Ly$\alpha$ production \citep{2025ApJ...985L...8U}. This suggests that AzTEC14 represents a more mature halo in which gas accretion is reduced \textcolor{blue}{at the center} but may still occur in the outskirts. %
In contrast, the AzTEC1 group overlaps with Ly$\alpha$-bright filaments and contains no QGs, indicating a less evolved state. Its members (e.g., A1 and A7) are predominantly in Stage~II, consistent with ongoing gas accretion sustaining active star formation. The Stage~I galaxy A5 is located in the outermost region, suggesting an early phase in which gas has accumulated but star formation has not yet fully ignited.

The three QGs are concentrated in the core of the most massive structure, suggesting that environmental effects play a role in quenching. 
As discussed in \citet{2025ApJ...985L...8U}, the cessation of gas accretion is likely a key factor. 
Without a fresh gas supply, galaxies rapidly consume their remaining reservoirs, leading to quenching. 
\textcolor{blue}{Alternatively, environmental processes such as virial heating, AGN feedback, may suppress the formation of molecular gas from newly supplied material, thereby limiting the replenishment of the cold molecular gas reservoirs required for sustained star formation.}
The relatively sparse population of transition objects in the $\Sigma_\star$–$\mu_\mathrm{gas}$ diagram may reflect the short timescale of this phase.
\textcolor{blue}{Together with morphological quenching (e.g., \cite{2009ApJ...707..250M})}, multiple quenching pathways may coexist, with their relative importance depending on the physical conditions of individual systems.

The ADF22 system, including both DSFGs and QGs, provides a case study of galaxy evolution within a proto-cluster core. More broadly, \textcolor{blue}{cool} gas accretion from the cosmic web and its eventual suppression by hot halo gas are thought to be fundamental processes regulating galaxy evolution at cosmic noon and earlier epochs. The evolutionary sequence identified in ADF22, therefore, offers a valuable framework for understanding the formation and transformation of massive galaxies in dense environments.
This framework links galaxy-scale transformation to the large-scale cosmic web, suggesting that filamentary gas supply and its eventual suppression govern both star formation and structural evolution.












\section{Conclusions}

We present a comprehensive census of dust and molecular gas in 18 DSFGs at $z=3.09$ embedded in Ly$\alpha$-traced cosmic web filaments in the SSA22 proto-cluster. 
Our main results are summarized as follows:

\begin{enumerate}

\item Using multi-band ALMA photometry combined with Herschel/SPIRE data, we construct well-sampled far-infrared SEDs and find that the DSFGs span $L_{\rm FIR}\sim10^{11}$--$10^{13}\,L_\odot$,  with median values of $\log(L_{\rm FIR}/L_\odot)=11.56^{+0.32}_{-0.37}$ and $T_{\rm d}=25^{+6}_{-3}$~K. 
This indicates that a substantial fraction of star formation in these systems occurs in relatively cold dust environments.

\item High-resolution JWST and ALMA imaging for a subset of eight DSFGs reveals a positive correlation between stellar and dust surface densities, suggesting that structural growth of the stellar component proceeds in concert with the interstellar medium.

\item We detect CO emission in multiple transitions (12 in CO(1--0), 18 in CO(3--2), and 4 in CO(8--7)). 
The median brightness temperature ratio $r_{31}=0.66_{-0.04}^{+0.05}$ is consistent with field galaxies, indicating that the bulk molecular gas is in a moderately excited state. 

\item \textcolor{blue}{High-$J$ CO lines show relatively low excitation, and the CO(8--7)/CO(3--2) ratio correlates with the star-formation rate surface density. This trend is broadly consistent with a scenario in which the molecular gas excitation is primarily regulated by star-formation activity and remains moderate in galaxies with relatively extended star-forming regions.}

\item \textcolor{blue}{Comparison between dust-based gas masses and CO luminosities suggests \textcolor{magenta}{that a broad range of CO-to-molecular gas conversion factors, $\alpha_{\rm CO}\sim1$--3.6, may be plausible, although the uncertainty in the GDR remains significant.} This implies that the molecular gas properties of ADF22 DSFGs are not necessarily consistent with those of extreme compact starbursts.}

\item The molecular gas fraction and depletion time show a clear stellar-mass dependence: lower-mass galaxies exhibit more pronounced excesses relative to field scaling relations, while more massive systems are broadly consistent with the field. 
These results suggest that proto-cluster galaxies 
may evolve under different balances between gas supply and gas consumption, with the observed stellar-mass dependence likely reflecting different evolutionary stages.

\item \textcolor{blue}{By combining DSFGs with quiescent galaxies in the same field, we find an observationally motivated sequence from gas-rich systems with low stellar surface densities to compact, gas-poor galaxies with high stellar surface densities. 
The distribution in the $\Sigma_\star$–$\mu_\mathrm{gas}$ plane suggests that gas depletion and structural transformation may proceed together during massive galaxy evolution.}

\end{enumerate}

\textcolor{blue}{Taken together, these results support a picture in which massive galaxy evolution in dense environments is shaped by the interplay between gas supply, star formation, and structural transformation within the cosmic web environment. 
The DSFG population likely traces a gas-rich growth phase that may be sustained, at least in part, by continued gas supply along cosmic web filaments, while gas depletion and increasing central concentration are associated with the emergence of compact quiescent galaxies. 
These results link galaxy-scale evolution to the surrounding large-scale structure and provide a coherent framework for massive galaxy formation in proto-cluster environments.}

\begin{table*}
\caption{Dust and stellar properties of the ADF22 DSFGs}
\label{tab:dust_stellar_properties}
\centering

\begin{tabular}{lccccccccc}
\hline

ID &
$M_\star$ &
SFR &
$L_{\rm FIR}$ &
$M_{\rm dust}$ &
$T_{\rm dust}$ &
$T_{\rm peak}$ &
$\beta_{\rm IR}$ &
$R_{\rm e,dust}$ &
$R_{\rm e,\star}$ \\

&
$\log(M_\odot)$ &
$(M_\odot\,\mathrm{yr}^{-1})$ &
$\log(L_\odot)$ &
$\log(M_\odot)$ &
(K) &
(K) &
&
(kpc) &
(kpc) \\

\hline

ADF22.A1 & $11.19^{+0.24}_{-0.29}$ & $551^{+351}_{-171}$ & $12.58^{+0.06}_{-0.06}$ & $9.36^{+0.04}_{-0.04}$ & $26.6^{+1.5}_{-1.4}$ & $26.9^{+1.1}_{-1.1}$ & $2.1^{+0.1}_{-0.1}$ & $3.05^{+0.37}_{-0.37}$ & $4.17^{+0.48}_{-0.48}$ \\
ADF22.A3 & $10.48^{+0.21}_{-0.27}$ & $145^{+42}_{-29}$ & $11.74^{+0.15}_{-0.16}$ & $9.18^{+0.12}_{-0.10}$ & $21.1^{+2.6}_{-2.1}$ & $21.8^{+2.0}_{-1.8}$ & $2.1^{+0.2}_{-0.2}$ & $0.99^{+0.13}_{-0.13}$ & $1.94^{+0.23}_{-0.23}$ \\
ADF22.A4 & $10.90^{+0.24}_{-0.32}$ & $248^{+160}_{-91}$ & $11.72^{+0.16}_{-0.15}$ & $9.11^{+0.09}_{-0.07}$ & $21.9^{+2.7}_{-2.1}$ & $22.1^{+2.0}_{-1.6}$ & $2.0^{+0.2}_{-0.2}$ & $1.00^{+0.15}_{-0.15}$ & $1.68^{+0.21}_{-0.21}$ \\
ADF22.A5 & $10.65^{+0.23}_{-0.27}$ & $205^{+131}_{-75}$ & $12.74^{+0.04}_{-0.05}$ & $8.77^{+0.04}_{-0.04}$ & $36.3^{+2.1}_{-2.0}$ & $35.9^{+1.5}_{-1.4}$ & $1.9^{+0.1}_{-0.1}$ & $1.26^{+0.17}_{-0.17}$ & $1.83^{+0.22}_{-0.22}$ \\
ADF22.A6 & $10.66^{+0.16}_{-0.23}$ & $140^{+50}_{-40}$ & $11.83^{+0.16}_{-0.16}$ & $8.83^{+0.11}_{-0.10}$ & $24.1^{+2.9}_{-2.5}$ & $24.8^{+2.3}_{-2.1}$ & $2.2^{+0.2}_{-0.2}$ & $0.97^{+0.13}_{-0.13}$ & $1.25^{+0.17}_{-0.17}$ \\
ADF22.A7 & $10.64^{+0.20}_{-0.36}$ & $146^{+39}_{-34}$ & $12.00^{+0.15}_{-0.15}$ & $9.00^{+0.08}_{-0.10}$ & $25.4^{+2.6}_{-2.3}$ & $25.2^{+2.0}_{-1.9}$ & $2.0^{+0.1}_{-0.1}$ & $1.05^{+0.12}_{-0.12}$ & $1.65^{+0.21}_{-0.21}$ \\
ADF22.A8 & $10.94^{+0.14}_{-0.17}$ & $87^{+44}_{-31}$ & $11.60^{+0.22}_{-0.24}$ & $8.84^{+0.15}_{-0.14}$ & $24.7^{+4.1}_{-3.4}$ & $23.8^{+3.3}_{-2.8}$ & $1.8^{+0.2}_{-0.2}$ & --- & $^\dagger0.88^{+0.19}_{-0.19}$ \\
ADF22.A9 & $11.19^{+0.11}_{-0.13}$ & $226^{+103}_{-81}$ & $11.52^{+0.25}_{-0.26}$ & $8.51^{+0.19}_{-0.18}$ & $25.3^{+3.8}_{-3.3}$ & $25.3^{+3.9}_{-3.3}$ & $2.0^{\rm fixed}$ & --- & --- \\
ADF22.A10 & $10.08^{+0.30}_{-0.30}$ & $66^{+52}_{-26}$ & $11.46^{+0.25}_{-0.26}$ & $8.57^{+0.13}_{-0.12}$ & $24.5^{+4.2}_{-3.4}$ & $24.4^{+3.5}_{-2.9}$ & $2.0^{+0.2}_{-0.2}$ & $0.94^{+0.17}_{-0.17}$ & $1.76^{+0.28}_{-0.28}$ \\
ADF22.A11 & $10.89^{+0.17}_{-0.24}$ & $142^{+98}_{-67}$ & $11.15^{+0.20}_{-0.19}$ & $8.72^{+0.13}_{-0.11}$ & $19.1^{+2.5}_{-2.1}$ & $20.4^{+2.1}_{-1.8}$ & $2.4^{+0.2}_{-0.2}$ & $1.06^{+0.30}_{-0.30}$ & $2.08^{+0.32}_{-0.32}$ \\
ADF22.A12 & $10.90^{+0.10}_{-0.11}$ & $54^{+39}_{-19}$ & $11.28^{+0.27}_{-0.28}$ & $8.52^{+0.20}_{-0.20}$ & $23.3^{+3.7}_{-3.1}$ & $23.3^{+3.7}_{-3.1}$ & $2.0^{\rm fixed}$ & --- & $^{\dagger\dagger}2.21^{+0.37}_{-0.37}$ \\
ADF22.A13 & $10.72^{+0.13}_{-0.17}$ & $74^{+46}_{-28}$ & $11.61^{+0.25}_{-0.29}$ & $8.48^{+0.19}_{-0.18}$ & $26.3^{+4.0}_{-3.6}$ & $26.3^{+4.0}_{-3.6}$ & $2.0^{\rm fixed}$ & --- & $^\dagger1.53^{+0.25}_{-0.25}$ \\
ADF22.A14 & $10.53^{+0.24}_{-0.33}$ & $84^{+58}_{-34}$ & $11.34^{+0.26}_{-0.30}$ & $8.71^{+0.21}_{-0.19}$ & $22.6^{+3.3}_{-3.0}$ & $22.6^{+3.3}_{-3.0}$ & $2.0^{\rm fixed}$ & --- & $^{\dagger\dagger}2.79^{+0.44}_{-0.44}$ \\
ADF22.A15 & $10.66^{+0.18}_{-0.26}$ & $79^{+48}_{-33}$ & $11.80^{+0.24}_{-0.31}$ & $7.90^{+0.17}_{-0.15}$ & $34.1^{+6.8}_{-5.9}$ & $34.4^{+5.9}_{-5.4}$ & $2.0^{+0.2}_{-0.2}$ & --- & $0.34^{+0.06}_{-0.06}$ \\
ADF22.A16 & $10.38^{+0.19}_{-0.27}$ & $63^{+36}_{-22}$ & $11.11^{+0.40}_{-0.41}$ & $8.26^{+0.22}_{-0.21}$ & $24.6^{+6.6}_{-4.7}$ & $24.3^{+5.8}_{-4.3}$ & $1.9^{+0.2}_{-0.2}$ & --- & $2.52^{+0.39}_{-0.39}$ \\
ADF22.A17 & $10.32^{+0.27}_{-0.32}$ & $67^{+44}_{-24}$ & $11.08^{+0.41}_{-0.48}$ & $8.66^{+0.27}_{-0.23}$ & $21.1^{+4.5}_{-3.7}$ & $21.1^{+4.5}_{-3.7}$ & $2.0^{\rm fixed}$ & --- & $1.35^{+0.16}_{-0.16}$ \\
ADF22.A38 & $10.40^{+0.31}_{-0.32}$ & $57^{+32}_{-22}$ & $11.20^{+0.51}_{-0.73}$ & $8.23^{+0.38}_{-0.28}$ & $25.1^{+7.5}_{-6.7}$ & $25.1^{+7.5}_{-6.7}$ & $2.0^{\rm fixed}$ & --- & $1.26^{+0.12}_{-0.12}$ \\
ADF22.A41 & $10.08^{+0.10}_{-0.18}$ & $67^{+32}_{-19}$ & $11.32^{+0.48}_{-0.62}$ & $7.94^{+0.32}_{-0.25}$ & $28.8^{+8.4}_{-7.4}$ & $28.8^{+8.4}_{-7.4}$ & $2.0^{\rm fixed}$ & --- & --- \\

\hline
\end{tabular}

\begin{flushleft}
\footnotesize
Note.
Object IDs are identical to those used in \citet{2017ApJ...835...98U,2018PASJ...70...65U,2019Sci...366...97U,2025A&A...699A.324H}. 
The far-infrared luminosity, $L_{\rm FIR}$, defined over the rest-frame wavelength range of 42.5--122.5\,\micron, as well as the dust temperature ($T_{\rm dust}$), peak temperature ($T_{\rm peak}$), and dust emissivity index ($\beta_{\rm IR}$), are consistently derived from a single dust SED fitting analysis performed in this work. 
The stellar masses and star formation rates (SFRs) are adopted from the SED fitting results of \citet{2025A&A...699A.324H}. 
All effective radii are circularized values. 
The effective radii at 870\,$\mu$m and in the F444W band are adopted from \citet{2026ApJ...997...79U}. 
Symbols $\dagger$ and $\dagger\dagger$ indicate that the stellar sizes are measured from the F560W and F770W bands, respectively. 
ADF22.A9 is a QSO \citep{1998ApJ...492..428S,2015ApJ...815L...8U}. 
ADF22.A41 shows a possible spatial offset between the dust and stellar components (Fig.~\ref{fig:finding_chart}).
\end{flushleft}

\end{table*}

\begin{table*}
\caption{CO(3--2) and CO(1--0) measurements of the ADF22 galaxies}
\label{tab:co32co10}
\centering

\begin{tabular}{lccccccc}
\hline

ID &
$z_{\rm spec}$ &
$\sigma_{\rm CO(3-2)}$ &
$S_{\rm CO(3-2)}\Delta v$ &
$\log L'_{\rm CO(3-2)}$ &
$S_{\rm CO(1-0)}\Delta v$ &
$\log L'_{\rm CO(1-0)}$ &
$r_{31}$ \\

&
&
(km\,s$^{-1}$) &
(Jy\,km\,s$^{-1}$) &
(K\,km\,s$^{-1}$\,pc$^2$) &
(Jy\,km\,s$^{-1}$) &
(K\,km\,s$^{-1}$\,pc$^2$) &
\\

\hline

ADF22.A1 & $3.09083^{+0.00141}_{-0.00152}$ & $363^{+224}_{-80}$ & $1.49\pm0.04$ & $10.84^{+0.01}_{-0.01}$ & $0.19\pm0.01$ & $10.89^{+0.03}_{-0.03}$ & $0.89^{+0.08}_{-0.07}$ \\
ADF22.A3 & $3.09786^{+0.00079}_{-0.00078}$ & $364^{+112}_{-63}$ & $1.08\pm0.13$ & $10.70^{+0.05}_{-0.06}$ & $0.13\pm0.03$ & $10.75^{+0.10}_{-0.13}$ & $0.90^{+0.33}_{-0.21}$ \\
ADF22.A4 & $3.09154^{+0.00003}_{-0.00003}$ & $95^{+2}_{-2}$ & $0.52\pm0.01$ & $10.38^{+0.01}_{-0.01}$ & $0.10\pm0.01$ & $10.61^{+0.03}_{-0.03}$ & $0.59^{+0.05}_{-0.04}$ \\
ADF22.A5 & $3.08942^{+0.00014}_{-0.00015}$ & $249^{+8}_{-6}$ & $1.12\pm0.05$ & $10.71^{+0.02}_{-0.02}$ & $0.19\pm0.01$ & $10.89^{+0.03}_{-0.04}$ & $0.66^{+0.06}_{-0.05}$ \\
ADF22.A6 & $3.09660^{+0.00082}_{-0.00072}$ & $327^{+69}_{-45}$ & $0.50\pm0.06$ & $10.37^{+0.05}_{-0.06}$ & $<0.06$ & $<10.43$ & $>0.87$ \\
ADF22.A7 & $3.09536^{+0.00014}_{-0.00014}$ & $258^{+5}_{-5}$ & $0.64\pm0.03$ & $10.47^{+0.02}_{-0.02}$ & $0.11\pm0.01$ & $10.66^{+0.03}_{-0.03}$ & $0.65^{+0.05}_{-0.05}$ \\
ADF22.A8 & $3.09009^{+0.00135}_{-0.00128}$ & $471^{+158}_{-105}$ & $0.23\pm0.05$ & $10.03^{+0.09}_{-0.11}$ & --- & --- & --- \\
ADF22.A9 & $3.09579^{+0.00035}_{-0.00032}$ & $166^{+16}_{-18}$ & $0.37\pm0.03$ & $10.24^{+0.04}_{-0.04}$ & $0.07\pm0.00$ & $10.43^{+0.03}_{-0.03}$ & $0.63^{+0.07}_{-0.07}$ \\
ADF22.A10 & $3.09260^{+0.00021}_{-0.00018}$ & $208^{+9}_{-8}$ & $0.36\pm0.03$ & $10.22^{+0.03}_{-0.04}$ & $0.06\pm0.02$ & $10.41^{+0.10}_{-0.13}$ & $0.65^{+0.24}_{-0.14}$ \\
ADF22.A11 & $3.09807^{+0.00080}_{-0.00168}$ & $229^{+198}_{-62}$ & $0.32\pm0.11$ & $10.17^{+0.13}_{-0.18}$ & $0.06\pm0.02$ & $10.41^{+0.14}_{-0.20}$ & $0.57^{+0.40}_{-0.23}$ \\
ADF22.A12 & $3.09734^{+0.00242}_{-0.00213}$ & $387^{+131}_{-146}$ & $0.29\pm0.09$ & $10.12^{+0.11}_{-0.15}$ & $<0.11$ & $<10.65$ & $>0.29$ \\
ADF22.A13 & $3.09104^{+0.00038}_{-0.00032}$ & $99^{+21}_{-23}$ & $0.24\pm0.04$ & $10.04^{+0.06}_{-0.07}$ & $<0.08$ & $<10.51$ & $>0.34$ \\
ADF22.A14 & $3.09890^{+0.00184}_{-0.00199}$ & $324^{+222}_{-130}$ & $0.75\pm0.30$ & $10.54^{+0.15}_{-0.22}$ & $0.12\pm0.06$ & $10.71^{+0.17}_{-0.28}$ & $0.68^{+0.65}_{-0.32}$ \\
ADF22.A15 & $3.08712^{+0.00008}_{-0.00008}$ & $80^{+5}_{-6}$ & $0.20\pm0.01$ & $9.97^{+0.03}_{-0.03}$ & $0.03\pm0.01$ & $10.07^{+0.15}_{-0.24}$ & $0.78^{+0.55}_{-0.23}$ \\
ADF22.A16 & $3.08908^{+0.00020}_{-0.00021}$ & $143^{+8}_{-8}$ & $0.14\pm0.01$ & $9.81^{+0.04}_{-0.04}$ & $0.04\pm0.01$ & $10.20^{+0.12}_{-0.18}$ & $0.41^{+0.20}_{-0.11}$ \\
ADF22.A17 & $3.09417^{+0.00025}_{-0.00023}$ & $196^{+8}_{-9}$ & $0.30\pm0.02$ & $10.14^{+0.03}_{-0.03}$ & $0.06\pm0.02$ & $10.39^{+0.15}_{-0.22}$ & $0.55^{+0.36}_{-0.16}$ \\
ADF22.A38 & $3.08815^{+0.00028}_{-0.00026}$ & $121^{+24}_{-16}$ & $0.21\pm0.02$ & $9.98^{+0.04}_{-0.05}$ & $<0.07$ & $<10.45$ & $>0.34$ \\
ADF22.A41 & $3.09907^{+0.00041}_{-0.00049}$ & $114^{+28}_{-27}$ & $0.04\pm0.01$ & $9.26^{+0.06}_{-0.07}$ & $<0.03$ & $<10.16$ & $>0.12$ \\

\hline
\end{tabular}

\begin{flushleft}
\footnotesize
Note.
The spectroscopic redshift ($z_{\rm spec}$) and velocity dispersion ($\sigma_{\rm CO(3-2)}$) are derived from the CO(3--2) spectra. 
The line luminosities are given as $\log L'$ in units of
K\,km\,s$^{-1}$\,pc$^2$.
\end{flushleft}

\end{table*}

\begin{table*}
\caption{High-$J$ CO measurements of ADF22.A1, A4, A6, and A7}
\label{tab:highj_co}
\centering

\begin{tabular}{lcccc}
\hline

ID &
$S_{\rm CO(8-7)}\Delta v$ &
$S_{\rm CO(9-8)}\Delta v$ &
$S_{\rm CO(12-11)}\Delta v$ &
$S_{\rm CO(8-7)}/S_{\rm CO(3-2)}$ \\

&
(Jy\,km\,s$^{-1}$) &
(Jy\,km\,s$^{-1}$) &
(Jy\,km\,s$^{-1}$) &
\\

\hline

ADF22.A1 & $0.66\pm0.11$ & $0.58\pm0.10$ & $<1.47$ & $0.44\pm0.08$ \\
ADF22.A4 & $0.61\pm0.06$ & $0.40\pm0.06$ & $0.26\pm0.04$ & $1.16\pm0.12$ \\
ADF22.A6 & $0.40\pm0.08$ & $0.43\pm0.10$ & $<1.38$ & $0.79\pm0.18$ \\
ADF22.A7 & $0.38\pm0.07$ & $0.29\pm0.08$ & $<1.25$ & $0.59\pm0.11$ \\

\hline
\end{tabular}

\begin{flushleft}
\footnotesize
Note.
The CO(9--8) detection of ADF22.A4 was first reported by \citet{2017PASJ...69...45H}. Our measured line flux is consistent with their reported value.
\end{flushleft}

\end{table*}

\begin{table*}
\caption{Molecular gas properties}
\label{tab:molgas}
\centering

\begin{tabular}{lcccccc}
\hline

ID & $\log M_{\rm mol}$ & $\log \mu_{\rm gas}$ & $\log \tau_{\rm dep}$ & $\log \Sigma_\star$ & Stage & AGN \\

& (${\rm M_\odot}$) & & (yr) & (${\rm M_\odot\,kpc^{-2}}$) & & \\

\hline
ADF22.A1 & $11.44^{+0.03}_{-0.03}$ & $0.25^{+0.24}_{-0.29}$ & $8.70^{+0.22}_{-0.17}$ & $9.15^{+0.25}_{-0.33}$ & II & Y \\
ADF22.A3 & $11.30^{+0.10}_{-0.13}$ & $0.82^{+0.23}_{-0.33}$ & $9.14^{+0.14}_{-0.17}$ & $9.10^{+0.22}_{-0.32}$ & I & N \\
ADF22.A4 & $11.17^{+0.03}_{-0.03}$ & $0.27^{+0.24}_{-0.32}$ & $8.78^{+0.22}_{-0.20}$ & $9.65^{+0.25}_{-0.37}$ & II & Y \\
ADF22.A5 & $11.45^{+0.03}_{-0.04}$ & $0.79^{+0.23}_{-0.28}$ & $9.13^{+0.22}_{-0.20}$ & $9.33^{+0.24}_{-0.32}$ & I & N \\
ADF22.A6 & $11.10^{+0.06}_{-0.06}$ & $0.44^{+0.16}_{-0.25}$ & $8.96^{+0.14}_{-0.17}$ & $9.67^{+0.18}_{-0.30}$ & II & Y \\
ADF22.A7 & $11.21^{+0.03}_{-0.03}$ & $0.57^{+0.20}_{-0.37}$ & $9.05^{+0.11}_{-0.12}$ & $9.41^{+0.22}_{-0.42}$ & II & Y \\
ADF22.A8 & $10.77^{+0.09}_{-0.11}$ & $-0.17^{+0.16}_{-0.22}$ & $8.83^{+0.19}_{-0.24}$ & $10.26^{+0.20}_{-0.34}$ & III & N \\
ADF22.A9 & $10.99^{+0.03}_{-0.03}$ & --- & --- & --- & --- & Y \\
ADF22.A10 & $10.96^{+0.10}_{-0.13}$ & $0.88^{+0.31}_{-0.36}$ & $9.14^{+0.26}_{-0.28}$ & $8.79^{+0.31}_{-0.39}$ & I & N \\
ADF22.A11 & $10.97^{+0.14}_{-0.20}$ & $0.08^{+0.20}_{-0.36}$ & $8.82^{+0.25}_{-0.40}$ & $9.45^{+0.19}_{-0.33}$ & III & N \\
ADF22.A12 & $10.86^{+0.12}_{-0.15}$ & $-0.04^{+0.15}_{-0.20}$ & $9.13^{+0.25}_{-0.27}$ & $9.41^{+0.15}_{-0.23}$ & III & Y \\
ADF22.A13 & $10.78^{+0.07}_{-0.08}$ & $0.06^{+0.14}_{-0.19}$ & $8.91^{+0.22}_{-0.23}$ & $9.56^{+0.17}_{-0.27}$ & III & N \\
ADF22.A14 & $11.26^{+0.17}_{-0.28}$ & $0.73^{+0.27}_{-0.54}$ & $9.34^{+0.26}_{-0.43}$ & $8.84^{+0.26}_{-0.42}$ & I & N \\
ADF22.A15 & $10.63^{+0.15}_{-0.24}$ & $-0.03^{+0.22}_{-0.42}$ & $8.73^{+0.24}_{-0.39}$ & $10.81^{+0.21}_{-0.40}$ & III & N \\
ADF22.A16 & $10.75^{+0.12}_{-0.18}$ & $0.37^{+0.21}_{-0.36}$ & $8.95^{+0.22}_{-0.29}$ & $8.78^{+0.21}_{-0.35}$ & II & N \\
ADF22.A17 & $10.95^{+0.15}_{-0.22}$ & $0.63^{+0.29}_{-0.46}$ & $9.13^{+0.25}_{-0.33}$ & $8.96^{+0.28}_{-0.37}$ & I & N \\
ADF22.A38 & $10.72^{+0.05}_{-0.06}$ & $0.32^{+0.31}_{-0.33}$ & $8.96^{+0.20}_{-0.23}$ & $9.40^{+0.31}_{-0.35}$ & II & N \\
ADF22.A41 & $9.99^{+0.07}_{-0.08}$ & --- & --- & --- & --- & N \\
\hline
\end{tabular}

\begin{flushleft}
\footnotesize
Note.
The "Stage" column lists the evolutionary stages defined in Sec~5.2.
\end{flushleft}

\end{table*}

\begin{ack}
We thank the anonymous referee for constructive comments and suggestions.
We thank Dominik Riechers and Chelsea Sharon for sharing their JVLA calibrated data. 
We also acknowledge the significant contributions to ALMA program \#2021.1.00511.S by Fabrizio Arrigoni Battaia, Edith Falgarone, and Alba Vidal-Garcia.
This work is based on observations made with the NASA/ESA/CSA James Webb Space Telescope. The data were obtained from the Mikulski Archive for Space Telescopes at the Space Telescope Science Institute, which is operated by the Association of Universities for Research in Astronomy, Inc., under NASA contract NAS 5-03127 for JWST. These observations are associated with program \#3547 and \#6751.
This paper makes use of the following ALMA data: ADS/JAO.ALMA \#2013.1.00162.S, \#2015.1.00212.S, \#2016.1.00543.S, \# 2017.1.01343.S, \#2019.1.00313,  \#2021.1.00041.S, \#2021.1.00071.S, \#2021.1.00511, \#2022.1.00680.S, \#2022.1.00223.S, \#2023.1.01206. ALMA is a partnership of ESO (representing its member states), NSF (USA) and NINS (Japan), together with NRC (Canada), NSTC and ASIAA (Taiwan), and KASI (Republic of Korea), in cooperation with the Republic of Chile. The Joint ALMA Observatory is operated by ESO, AUI/NRAO and NAOJ.
The National Radio Astronomy Observatory is a facility of the National Science Foundation operated under cooperative agreement by Associated Universities, Inc. This work is based on the observations of Karl G. Jansky Very Large Array (VLA) (program ID: 16A-357; 21A-346). 
HU acknowledges support from JSPS KAKENHI Grant Numbers 23K20035, 25K01039.
HU is supported by the ALMA Japan Research Grant of NAOJ ALMA Project, NAOJ-ALMA-375 and JST FOREST Program Grant Number JP-MJFR256M.
\end{ack}

\appendix

\setcounter{figure}{0}
\renewcommand{\thefigure}{A\arabic{figure}}

\setcounter{table}{0}
\renewcommand{\thetable}{A\arabic{table}}

\section{Summary of JVLA Observations}\label{app:jvla_obs}

Table~\ref{table:obs} summarizes the JVLA observations and parameters of the obtained cubes.

\renewcommand{\arraystretch}{1} 
\begin{table*}
  \tbl{Summary of JVLA Observations}{%
  \begin{tabular}{cccccccccc}
  \hline\noalign{\vskip3pt}
Field & Program  & RA          & Dec.           & Date & Config. & $N_{\rm ant}$ & $t_{\rm int}$ & $\theta_{\rm beam}$   \\
        &                  & (J2000)  &   (J2000)     &         &              &                       &  (hr)              &   \\
  \hline\noalign{\vskip3pt}
A1A5A7A15 & 16A-357 & 22:17:33.2 & 00:17:18.0 & 2016 Feb 26 $-$ 2016 Apr 18 & C  & 27 & 7.31  &  3.34$^{\prime\prime}\times2.64^{\prime\prime}$ \\
       &         &            &            & 2016 May 04 $-$ 2016 May 16 & CnB& 27 & 6.48  &(pa= -10$^\circ$)                                                              \\
       & 21A-346 & 22:17:32.1 & 00:17:43.0 & 2021 Apr 17 $-$ 2021 May 25 & D  & 27 & 8.14  &                                                              \\
       &         &            &            & 2021 Jun 16 $-$ 2021 Jun 17 & C  & 27 & 2.39  &                                                              \\
A4A10A11A16A17   & 21A-346 & 22:17:36.9 & 00:18:19.4 & 2021 Apr 03 $-$ 2021 May 30 & D  & 27 & 9.31  &  3.27$^{\prime\prime}\times2.35^{\prime\prime}$ \\
       &         &            &            & 2021 Jun 25 $-$ 2021 Aug 31 & C  & 27 & 4.79  &       (pa= -14$^\circ$)                                                       \\
A3A6A9    & 21A-346 & 22:17:36.2 & 00:16:11.3 & 2021 Apr 11 $-$ 2021 May 27 & D  & 27 & 4.65  &  3.27$^{\prime\prime}\times2.62^{\prime\prime}$ \\
       &         &            &            & 2021 Jun 14 $-$ 2021 Aug 09 & C  & 27 & 8.38  &             (pa= -10$^\circ$)                                                   \\
        & 15B-329 &    22:17:35.2 & 00:15:38.0      & 2015 Nov 06 $-$ 2015 Dec 03 & D  & 22 $-$ 25 & 2.77 &                                                       \\
\hline\noalign{\vskip3pt}
A8A12A13A14    &  &  & &  mosaic of the all fields & &  &  &  3.22$^{\prime\prime}\times2.51^{\prime\prime}$ \\
 & & & & & & & &             (pa= -76$^\circ$)  \\
\hline\noalign{\vskip3pt}
  \end{tabular}}\label{table:catalog}
  \label{table:obs}
\end{table*}
\renewcommand{\arraystretch}{1}

\section{ALMA Continuum Images}\label{app:cont_images}

\begin{figure*}[!t]
  \begin{center}
\includegraphics[width=17cm]
{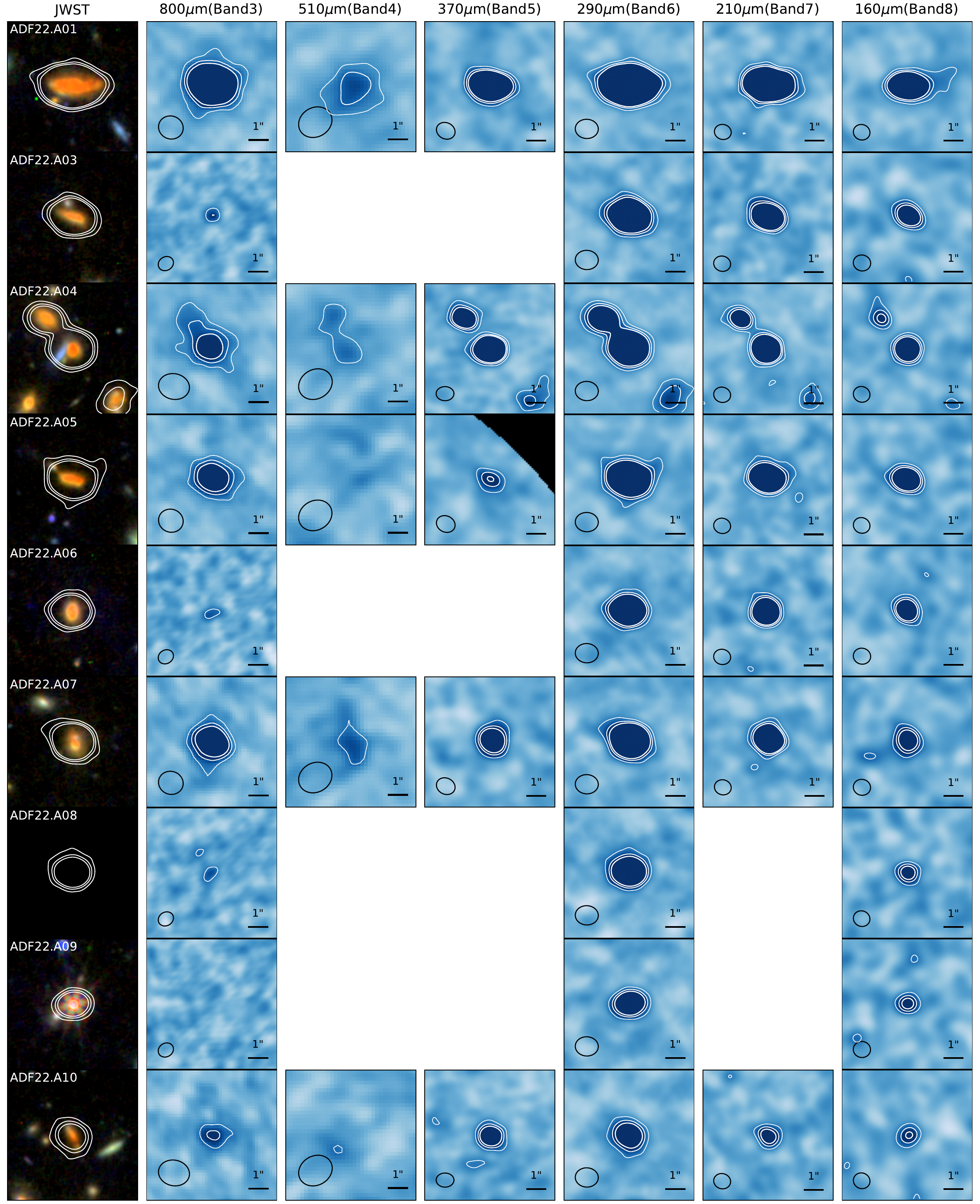}
  \end{center}
  \caption{%
Postage-stamp images of the 18 brightest DSFGs at $z=3.09$ in the ADF22 field.
As labeled, each source is shown with a JWST/NIRCam color composite image (F200W, F356W, and F444W) and ALMA continuum maps from Band~3 to Band~8. The synthesized beam is shown in the lower-left corner of each panel.
Each panel has a size of $7^{\prime\prime} \times 7^{\prime\prime}$. Contours are drawn at 3, 6, and 9$\sigma$; for the JWST images, the contours correspond to the ALMA Band~6 continuum emission.
Panels are left blank for bands in which a given source is not covered by ALMA observations.
Each DSFG has ALMA photometric measurements in two to six bands.
}%
  \label{fig:cont_maps1}
\end{figure*}

\begin{figure*}[!t]
  \begin{center}
\includegraphics[width=17cm]
{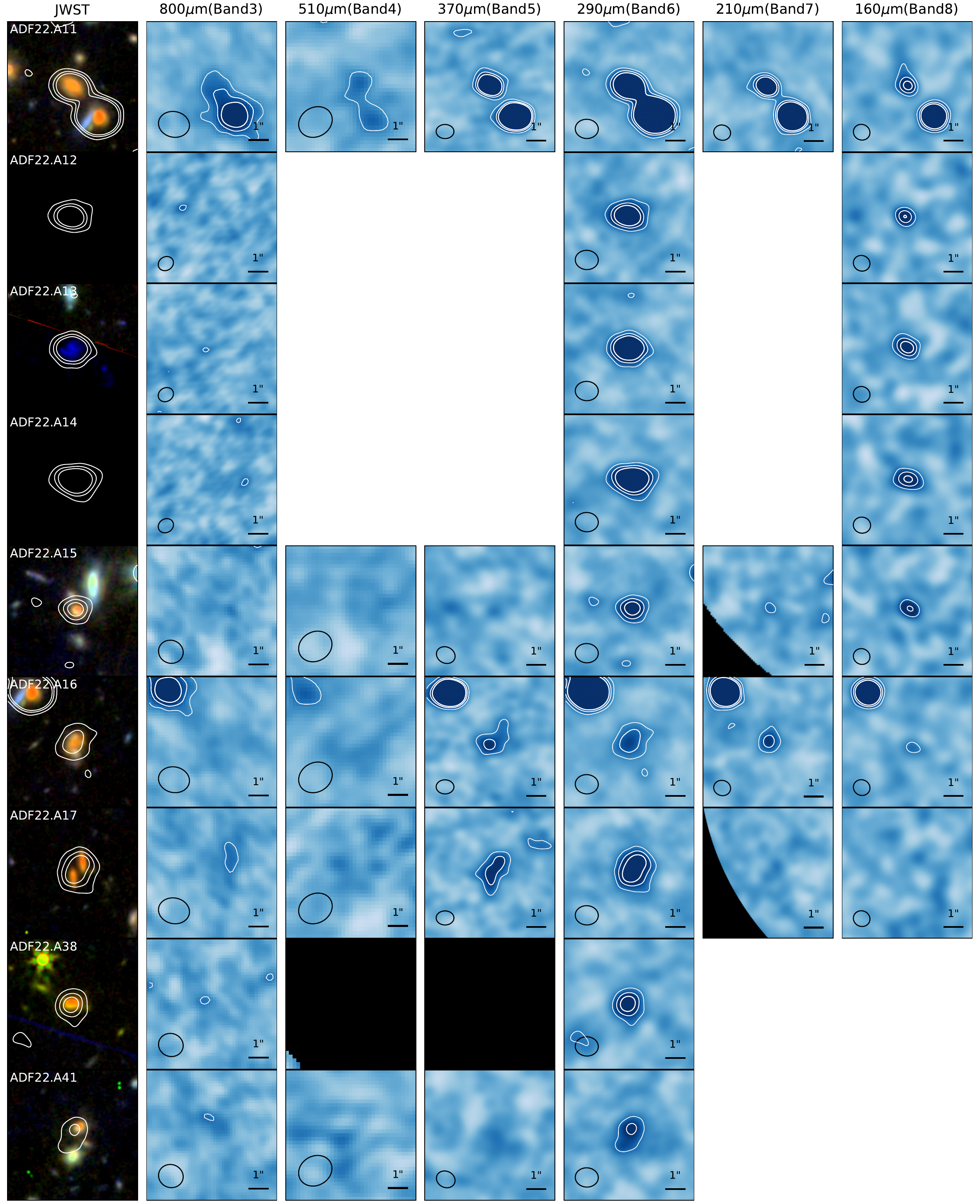}
  \end{center}
  \caption{%
Same as Fig.~2, but for ADF22.A11--ADF22.A41.
}%
  \label{fig:cont_maps2}
\end{figure*}

Figures~\ref{fig:cont_maps1} and \ref{fig:cont_maps2} present ALMA Band~3–8 continuum images of the 18 DSFGs in the ADF22 field. For reference, NIRCam color images are also shown with ALMA Band~6 contours as in Fig.~\ref{fig:finding_chart}.

\section{High-J CO spectra}\label{app:cont_images}

\begin{figure*}
  \begin{center}
\includegraphics[width=17cm]
{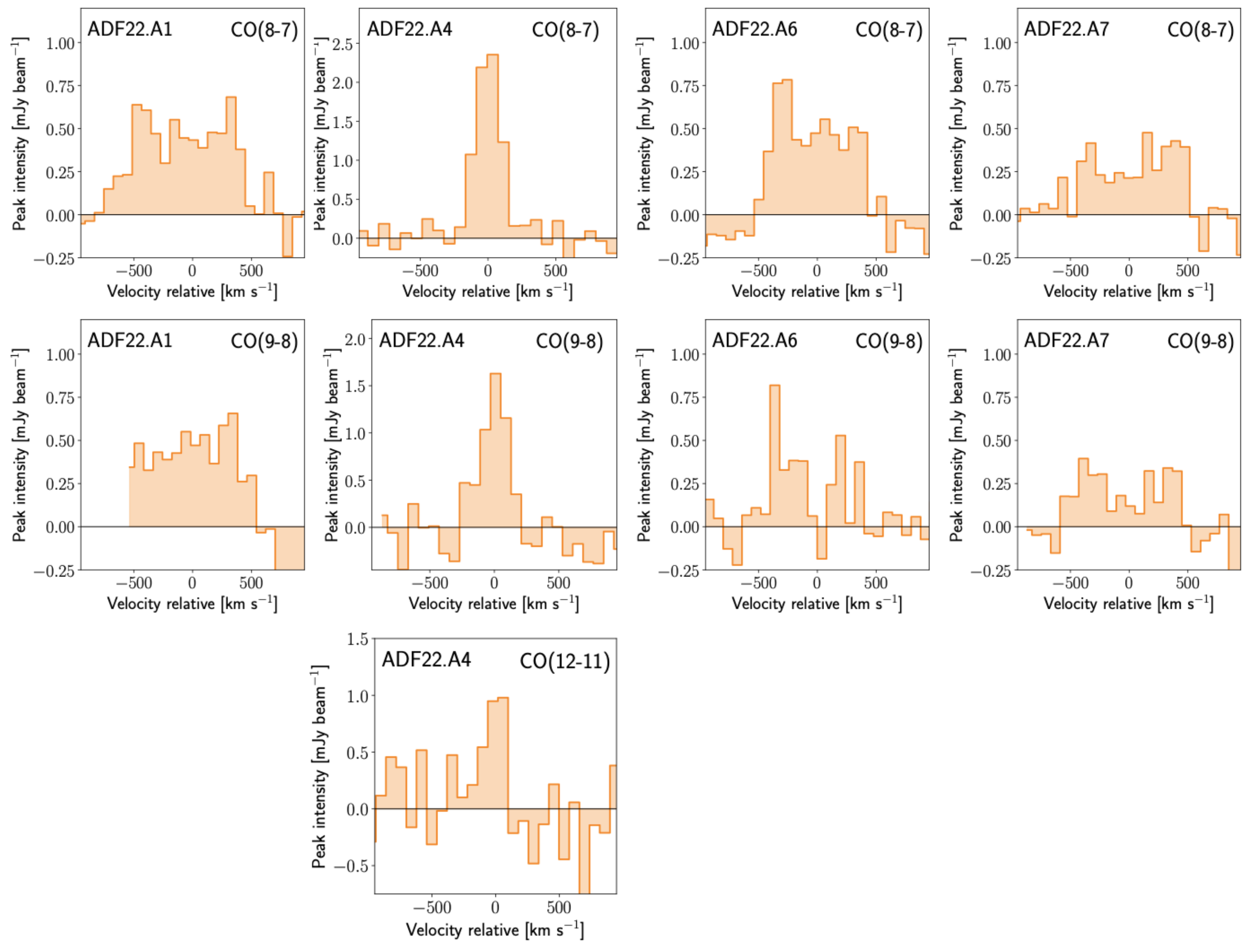}
  \end{center}
  \caption{%
High-$J$ CO spectra of the four sources ADF22.A1, A4, A6, and A7, extracted at the peak-intensity positions.
}%
  \label{fig:highjspec}
\end{figure*}

Figures~\ref{fig:highjspec} presents spectra of CO(8-7), CO(9-8), and CO(12-11) for detections.

\section{Submillimeter Photometry}
\label{app:submm_flux}

Table~\ref{tab:photometry} summarizes the Herschel/SPIRE and ALMA photometry of the 18 ADF22 DSFGs at $z_{\rm spec}\simeq3.09$.

\begin{table*}
\centering
\caption{ALMA and Herschel/SPIRE photometry of the ADF22 DSFGs.}
\label{tab:photometry}
\resizebox{\textwidth}{!}{
\begin{tabular}{lccccccccc}
\hline
ID & SPIRE 250 & SPIRE 350 & SPIRE 500 & Band~8 & Band~7 & Band~6 & Band~5 & Band~4 & Band~3\\
 & 60\,$\mu$m & 90\,$\mu$m & 120\,$\mu$m & 160\,$\mu$m & 210\,$\mu$m & 290\,$\mu$m & 370\,$\mu$m & 510\,$\mu$m & 800\,$\mu$m\\
\hline
ADF22.A1 & $<31$ & $29\pm4$ & $31\pm5$ & $22.2\pm0.9$ & $14.67\pm0.18$ & $6.15\pm0.06$ & $2.50\pm0.04$ & $0.78\pm0.06$ & $0.151\pm0.005$\\
ADF22.A3 & $<11$ & $<12$ & $<16$ & $6.9\pm0.5$ & $5.06\pm0.11$ & $2.23\pm0.04$ & --- & --- & $0.057\pm0.009$\\
ADF22.A4 & $<17$ & $<18$ & $<18$ & $5.7\pm0.4$ & $5.3\pm0.2$ & $2.21\pm0.03$ & $0.86\pm0.04$ & $0.36\pm0.07$ & $0.058\pm0.007$\\
ADF22.A5 & $24\pm4$ & $32\pm5$ & $28\pm5$ & $12.6\pm0.5$ & $6.70\pm0.18$ & $2.53\pm0.06$ & $1.16\pm0.17$ & $<0.66$ & $0.065\pm0.005$\\
ADF22.A6 & $<11$ & $<12$ & $<16$ & $5.7\pm0.4$ & $3.62\pm0.11$ & $1.43\pm0.04$ & --- & --- & $0.030\pm0.007$\\
ADF22.A7 & $<11$ & $8\pm4$ & $9\pm5$ & $7.7\pm0.6$ & $4.61\pm0.14$ & $2.03\pm0.04$ & $0.97\pm0.04$ & $0.38\pm0.10$ & $0.061\pm0.004$\\
ADF22.A8 & $<11$ & $<12$ & $<16$ & $3.9\pm0.5$ & --- & $1.14\pm0.04$ & --- & --- & $0.040\pm0.010$\\
ADF22.A9 & $<11$ & $<12$ & $<16$ & $2.7\pm0.4$ & --- & $0.69\pm0.03$ & --- & --- & $<0.063$\\
ADF22.A10 & $<11$ & $<12$ & $<16$ & $2.3\pm0.5$ & $1.62\pm0.17$ & $0.82\pm0.05$ & $0.32\pm0.02$ & $<0.31$ & $0.020\pm0.003$\\
ADF22.A11 & $<11$ & $<12$ & $<16$ & $1.7\pm0.4$ & $1.66\pm0.07$ & $0.85\pm0.03$ & $0.336\pm0.013$ & $<0.28$ & $0.012\pm0.003$\\
ADF22.A12 & $<11$ & $<12$ & $<16$ & $1.9\pm0.3$ & --- & $0.57\pm0.03$ & --- & --- & $<0.084$\\
ADF22.A13 & $<11$ & $<12$ & $<16$ & $3.1\pm0.5$ & --- & $0.71\pm0.04$ & --- & --- & $<0.071$\\
ADF22.A14 & $<11$ & $<12$ & $<16$ & $2.6\pm0.4$ & --- & $0.81\pm0.04$ & --- & --- & $<0.11$\\
ADF22.A15 & $<11$ & $<12$ & $<16$ & $2.2\pm0.4$ & $1.3\pm0.3$ & $0.33\pm0.04$ & $<0.45$ & $<0.36$ & $<0.014$\\
ADF22.A16 & $<11$ & $<12$ & $<16$ & $<1.7$ & $0.89\pm0.15$ & $0.41\pm0.06$ & $0.150\pm0.012$ & $<0.38$ & $<0.019$\\
ADF22.A17 & $<11$ & $<12$ & $<16$ & $<3.2$ & $<3.7$ & $0.68\pm0.05$ & $0.34\pm0.06$ & $<0.42$ & $<0.021$\\
ADF22.A38 & $<11$ & $<12$ & $<16$ & --- & --- & $0.37\pm0.04$ & --- & --- & $<0.021$\\
ADF22.A41 & $<11$ & $<12$ & $<16$ & --- & --- & $0.30\pm0.04$ & --- & --- & $<0.011$\\
\hline
\end{tabular}
}
\begin{flushleft}
\footnotesize
Note. Flux densities are given in mJy. The second header row gives approximate rest-frame wavelengths assuming $z=3.09$. Upper limits correspond to $3\sigma$ limits. 
\end{flushleft}

\end{table*}

\section{Projected separation versus line-of-sight velocity offset}
\label{Asec:mhalo}

\begin{figure}
  \begin{center}
\includegraphics[width=8.5cm]
{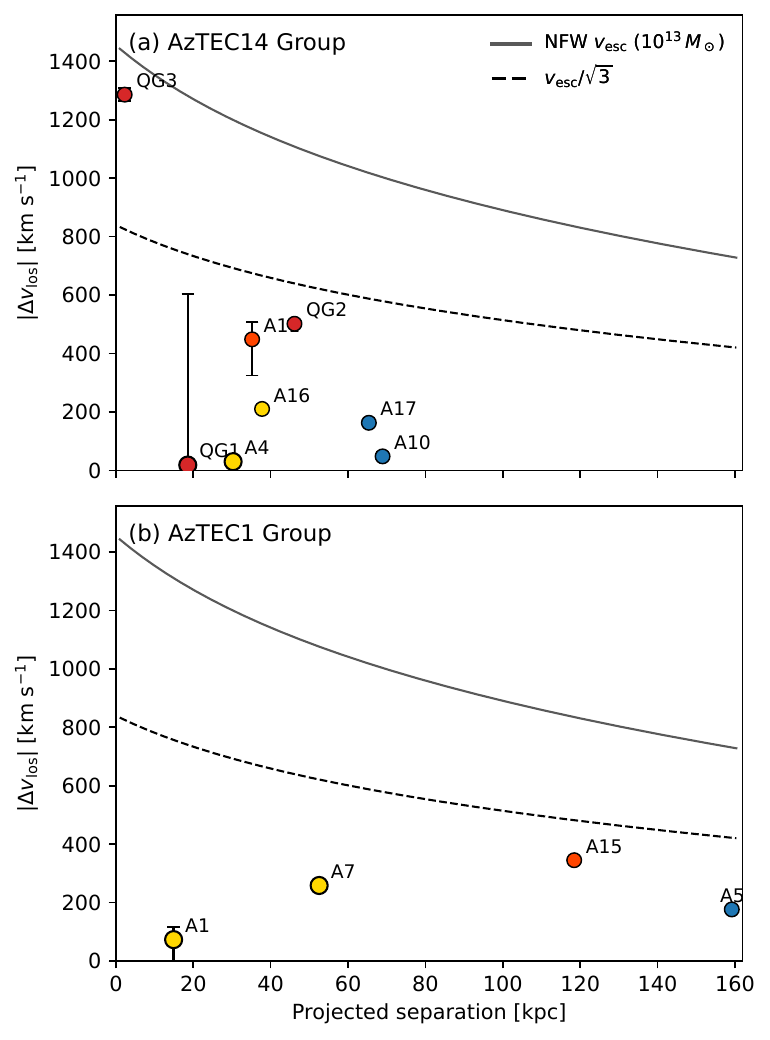}
  \end{center}
\caption{
Projected separation versus line-of-sight velocity offset for the DSFG groups in the ADF22 proto-cluster at $z\simeq3.09$.
The velocity offset is defined as $|\Delta v_{\rm los}| = |c(z - z_0)/(1+z_0)|$, where $z_0$ is the stellar-mass-weighted redshift of the two reference galaxies used to define the group center.
The projected separation is measured from the stellar-mass-weighted center of the two central galaxies: QG1 and A4 for the AzTEC14 group ({\it top}) and A1 and A7 for the AzTEC1 group ({\it bottom}).
Colors indicate the evolutionary stage of each galaxy (Stage I--Stage IV) \textcolor{blue}{as Fig.~\ref{fig:evolution}}.
Vertical error bars reflect the uncertainties in the spectroscopic redshifts.
The solid curve shows the escape velocity profile expected for an NFW halo with $M_{\rm vir}=10^{13}\,M_\odot$, while the dashed curve indicates the line-of-sight equivalent velocity scale $v_{\rm esc}/\sqrt{3}$ assuming isotropic velocities.
In both cases, the evolutionary stages look to be linked with the location. Relatively mature populations tend to reside close to the core.
{Alt text: Relation of distance and velocity dispersions.} 
}%
  \label{fig:halo}
\end{figure}

The velocity offset is defined as $|\Delta v_{\rm los}| = |c(z - z_0)/(1+z_0)|$, where $z_0$ is the stellar-mass-weighted redshift of the two most massive galaxies. The projected separation is measured from their stellar-mass-weighted center. The results are shown in Fig.~\ref{fig:halo}, together with escape velocity curves for an NFW halo of $10^{13}\,M_\odot$ (and a projected case scaled by $1/\sqrt{3}$), although the systems are not necessarily virialized. 

\bibliographystyle{apj}
\bibliography{reference_2024a}

\end{document}